\documentclass[reprint,superscriptaddress,amsmath,amssymb,aps,tightenlines,onecolumn,12pt]{revtex4-2}
\usepackage{graphicx}
\usepackage{newtxtext}
\usepackage{newtxmath}
\usepackage{natbib}
\usepackage{hyperref}
\usepackage{booktabs} 
\usepackage{overpic}
\usepackage[ruled,linesnumbered]{algorithm2e}
\hypersetup{
	colorlinks = true,
	hidelinks,
	urlcolor   = black,
	citecolor  = black,
}

\newcommand{\RomanNumeralCaps}[1]
\linenumbers

\newcommand{\bs}{\boldsymbol}

\newcommand{\mc}{\mathcal}

\newcommand\Rey{\mbox{\textit{Re}}}  % Reynolds number
\begin{document}

\title{Constructing wall turbulence using hierarchical hairpin vortices}

\author{Weiyu Shen}
\altaffiliation{These authors contributed equally to this work.}
\affiliation{State Key Laboratory for Turbulence and Complex Systems, School of Mechanics and Engineering Science, Peking University, Beijing 100871, China} 
\affiliation{Max Planck Institute for Solar System Research, 37077 Göttingen, Germany} 
\author{Yuchen Ge}
\altaffiliation{These authors contributed equally to this work.}
\affiliation{State Key Laboratory for Turbulence and Complex Systems, School of Mechanics and Engineering Science, Peking University, Beijing 100871, China}
\author{Zishuo Han}
\affiliation{State Key Laboratory for Turbulence and Complex Systems, School of Mechanics and Engineering Science, Peking University, Beijing 100871, China}
\author{Yaomin Zhao}
\email{Contact author: yaomin.zhao@pku.edu.cn}
\affiliation{State Key Laboratory for Turbulence and Complex Systems, School of Mechanics and Engineering Science, Peking University, Beijing 100871, China}
\affiliation{HEDPS-CAPT, Peking University, Beijing 100871, China}
\author{Yue Yang}
\email{Contact author: yyg@pku.edu.cn}
\affiliation{State Key Laboratory for Turbulence and Complex Systems, School of Mechanics and Engineering Science, Peking University, Beijing 100871, China}
\affiliation{HEDPS-CAPT, Peking University, Beijing 100871, China}

%\email[]{Your e-mail address}
%\homepage[]{Your web page}
%\thanks{}
%\altaffiliation{}
% \affiliation{}

%Collaboration name if desired (requires use of superscriptaddress
%option in \documentclass). \noaffiliation is required (may also be
%used with the \author command).
%\collaboration can be followed by \email, \homepage, \thanks as well.
%\collaboration{}
%\noaffiliation

\date{\today}

\begin{abstract}

Wall-bounded turbulence is characterized by coherent, worm-like structures such as hairpin vortices. The attached-eddy model provides a successful statistical framework for the log-law region, yet the complex geometry and multiscale nature of wall-turbulence vortices remain challenging for physics-based modelling. Here, we model wall turbulence as an ensemble of complex vortices, introducing a systematic approach to constructing turbulence fields enriched with hierarchically organized hairpin vortex packets. The geometry and organization of the vortex packets are calibrated to match observations, enabling the model to reproduce both attached and detached motions through a height-dependent core-size variation. Our model successfully reproduces the key statistical and structural features of wall turbulence, matching direct numerical simulations of turbulent channel flow at friction Reynolds numbers from 1,000 to 10,000. More importantly, it also reveals new insights into the coherent structures, emphasizing the role of vortex geometry, packet organization, and hierarchy in setting the attached/detached balance, meandering streaks and inclination angles, superstructure alignment, and the overall partition of contributions. Moreover, the constructed channel turbulence rapidly transitions into fully developed turbulence in direct numerical simulation, demonstrating its physical self-consistency and practical utility for initializing high-fidelity simulations. This approach significantly reduces computational costs associated with turbulence development while providing a flexible framework for testing and advancing turbulence models based on vortex structures.
\end{abstract}

\maketitle

\section{Introduction}\label{sec:intro}

Wall turbulence is ubiquitous, playing a critical role in a wide range of natural and engineering systems. 
A major aspect of understanding wall turbulence is identifying the important structures that govern its behaviour.
Wall-bounded turbulence exhibits a complex, multiscale nature characterized by hierarchical, interacting coherent structures governed by intricate vortex dynamics~\citep{Pope2000Turbulent}.
Building upon Townsend's attached-eddy hypothesis (AEH)~\citep{Townsend1976Structure}, hairpin vortices and their hierarchical organization have emerged as promising candidates for the fundamental structures that shape wall turbulence dynamics~\citep{Marusic2019Attached}. 
Hierarchical hairpin vortices arrange along the flow direction to form vortex packets~\citep{Adrian2007Hairpin} and superstructures~\citep{Kim1999Very,Lee2011Very,Baltzer2013Structural}, not only driving Reynolds-stress generation but also orchestrating the spatial distribution of energy across scales.

A substantial body of evidence from both experiments and simulations now supports the existence of self-similar wall-attached eddies in wall turbulence. Experimental investigations, employing techniques such as proper orthogonal decomposition and linear stochastic estimation, have revealed the presence of self-similar motions~\citep{Hellstrom2016Self,Baars2017Self,Chandran2017Two,Deshpande2020Two,hu2020wall,wang2022statistical}, while numerical studies based on coherent structure analysis have further confirmed their self-similarity and statistical characteristics~\citep{Lozano2012The,Hwang2018Wall,Hwang2020Statistical,Cheng2020Uncovering}.
By postulating hierarchical self-similar eddies that extend from the wall, the attached-eddy model (AEM) and its refined versions provide a theoretical framework that connects these coherent structures with the statistical properties of turbulence~\citep{Marusic2019Attached,Hu2023General}. 
The AEM effectively captures the dynamics of the logarithmic region and predicts key scaling behaviours, including the logarithmic scaling of streamwise and spanwise velocity moments, structure and correlation functions, logarithmic trends in two-point velocity statistics of different orders, and the streamwise $k_x^{-1}$ spectral law~\citep{Perry1995A,Woodcock2015Statistical,Yang2016Hierarchical,Baars2019Data1}. 
Despite these theoretical advances, controllable platforms for testing predictions based on instantaneous coherent structures and practical methodologies for embedding structure-resolved flow information into numerical simulations remain limited.

To bridge this gap, an alternative approach involves constructing turbulence itself to reproduce the three-dimensional (3D) instantaneous structures and statistical properties.
This inverse strategy not only enables the validation of various models and theories but also facilitates practical applications, such as generating realistic inlet or initial conditions for wall turbulence~\citep{Wu2017Inflow}. Generating such conditions is challenging because simple random noise perturbations lack physically consistent structures, can violate continuity, and are often suppressed by pressure-projection. Traditional synthetic methods, including the random methods~\citep{Kraichnan1970Diffusion,Smirnov2001Random}, digital filtering method~\citep{Klein2003A}, synthetic-eddy approaches~\citep{Jarrin2006A}, and volume forcing method~\citep{Spille2001Generation}, are widely used but often fail to reproduce the structural-statistical coherence of real turbulence. While these methods can generate pseudo-turbulence with targeted statistics, they require extensive adjustment zones for slow structural development, leading to high computational costs. This persistent gap between coherent structures and statistical fidelity underscores the need for novel physics-driven approaches to construct wall turbulence.

Coherent-structure-based modeling of wall-bounded turbulent flow seeks to bridge this gap by explicitly constructing flow fields. 
Early studies showed that hairpin vortices self-organize into streamwise-aligned packets~\citep{Zhou1999Mechanisms,Adrian2000Vortex}, providing a physical picture of the hierarchical arrangement of turbulence-producing structures and motivating the hairpin-packet model of \citet{Marusic2001On}, which helped link observed coherent motions with theoretical descriptions of wall turbulence. 
\citet{deSilva2016Uniform} generated 3D synthetic velocity fields using discrete attached vortices. Building on the hairpin-packet paradigm, they adopted packets of $\Lambda$-shaped vortices as representative eddies. Subsequent refinements to this strategy include incorporating spatial repulsion effects among same-scale vortices~\citep{deSilva2016Influence}, embedding streamwise meandering of large-scale structures~\citep{Eich2020Towards}, and developing spectral scaling-based AEM extensions~\citep{Chandran2020Spectral}.

Although these AEM-inspired synthetic efforts reproduce portions of the statistics of real turbulent flows, they still lack a comprehensive, readily tunable framework that can simultaneously capture both low- and higher-order statistics and recover key structural features across Reynolds numbers. 
With existing methods, flow fields are typically computed via the Biot–Savart law, based on idealized columnar $\Lambda$-shaped vortices that lack curvature or core-thickness variation. 
The rationale for adopting such highly simplified, non-adjustable vortex geometries is that the candidate eddy used in AEM is assumed as an average of many realizations of instantaneous structures, and that all eddy shapes tested in \citet{perry1986theoretical} give similar behaviour for the first- and second-order statistics. 
However, flow dynamics are known to be highly sensitive to the detailed geometry of representative structures~\citep{Deshpande2021Data}, and predictions beyond basic statistics, such as energy spectra and instantaneous vorticity fluctuation, explicitly depend on eddy shape \citep{Marusic2019Attached}.
Because methods for constructing complex near-wall vortices remain underdeveloped, these simplifications restrict the ability to replicate intricately structured vortices and to generate more realistic flow dynamics.
Moreover, in addition to traditional self-similar wall-attached eddies, wall turbulence contains wall-detached eddies \citep{Marusic2019Attached} and very-large-scale motions (VLSMs, i.e., superstructures) \citep{Kim1999Very,Lee2011Very,Deshpande2023Evidence}. While recent AEM extensions have incorporated representative eddies for both types \citep{Chandran2020Spectral}, the need to include these elements, however, requires further testing, especially with the more complex vortices considered in the present study.

Recent advances in numerical tools~\citep{Xiong2019Construction, Shen2023Role} now enable the bottom-up assembly of turbulence using intricately structured vortices as building blocks. This approach has successfully reproduced homogeneous isotropic turbulence without walls~\citep{Shen2024Designing}, offering fresh opportunities for modeling wall-bounded flows and achieving engineering goals~\citep{Subbareddy2006A}. 
In particular, differential-geometry and vortex-surface-based constructions permit explicit control of centreline geometry, local inclination, and core thickness, enabling complex vortical configurations within a single framework.
Building on these developments and moving beyond earlier AEM-based attempts, our goal is to offer a systematic framework for constructing 3D wall-bounded turbulence fields with complex-shaped coherent structures and accurate statistics, thereby providing a controllable platform for structure-based analysis, model validation, and mechanism testing.

This work introduces the synthetic wall-attached turbulence (SWAT) framework, a controllable platform for constructing wall turbulence from prescribed vortex organization. Complementing direct observations from simulations and experiments, this reverse assembly offers a transparent and flexible way to quantify the statistical contributions of coherent-structure geometry and hierarchy.
By prescribing hierarchically organised hairpin-vortex packets informed by AEM, SWAT enables systematic control of vortex geometry and organization, supporting validation of theoretical concepts and providing new insights into coherent motions.
Furthermore, SWAT also serves as a practical tool for generating initial conditions for direct numerical simulations (DNS) and large-eddy simulations (LES) at target Reynolds numbers without requiring additional data.
The remainder of the paper is organized as follows. \S 2 details the methods and configurations. \S 3 presents the results and statistical properties of SWAT across different Reynolds numbers. \S 4 exploits its tunable structural elements to yield physical insights into the geometry and organization of coherent structures. In \S 5, we construct channel flows, perform DNSs, and compare the computational cost against conventional methods. Finally, \S 6 provides discussions and concluding remarks.

\section{Methods}\label{sec:method} 

\subsection{Morphology of the hairpin vortex}\label{subsec:hairpin}

In the kinematic attached-eddy model, hierarchies of eddies are the fundamental building blocks of wall-bounded turbulent flows, with their shapes exhibiting self-similarity. Hence, designing hairpin vortices with uniform shapes but different scales is the first step in model establishment. 
Extensive numerical simulations and experimental observations \citep{Wu2017Transitional,Head1981New,Haidari1994The} indicate that real hairpin vortices possess complex centerline geometries and vortex core sizes that vary with height. 

\begin{figure}
	\centering
	\includegraphics[width=0.95\linewidth]{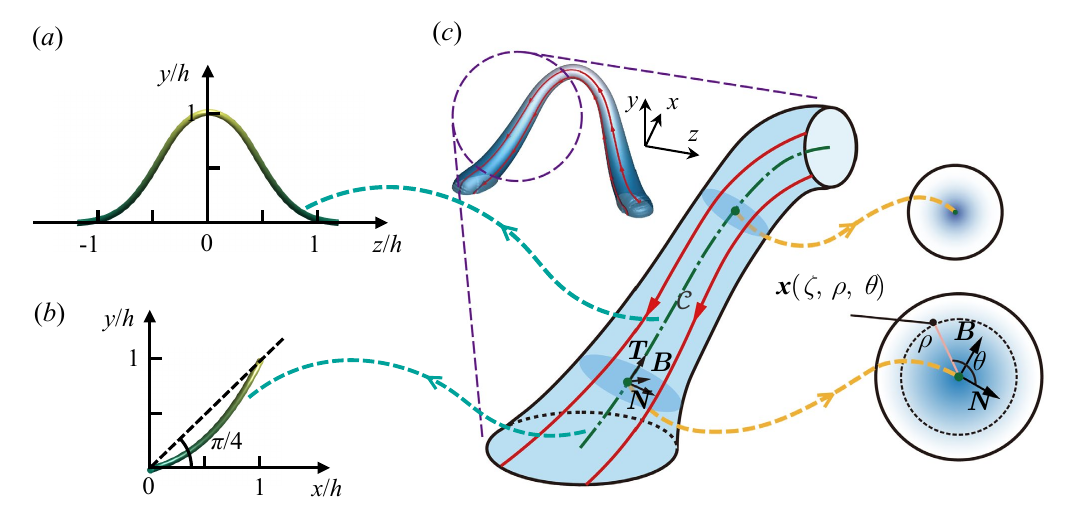}
	\caption{Geometry of a single hairpin vortex. (a) Front and (b) side views of the hairpin vortex centerline. (c) Vortex surface and a segment of a single hairpin vortex tube with variable thickness. The 3D hairpin vortex is visualized by VSF isosurface ($\phi_v=0.1$, blue) with embedded vortex lines (red solid). An enlarged schematic of the vortex tube segment is shown, where the vorticity is constructed in curved cylindrical coordinates $(\zeta,\rho,\theta)$. The vortex centerline $\mc C$ (green dash-dotted) is described in the Frenet--Serret frame ($ \bs{T} $, $ \bs{N} $, $ \bs{B} $). On each cross section of the vortex tube, the vorticity follows a Gaussian distribution with a continuously varying standard deviation $\sigma$.}
	\label{fig:hairpin}
\end{figure}

Thus, we specify the hairpin vortex centerline $\mc C$ as a parametric equation
\begin{equation}
	\boldsymbol{x}_c(\zeta)=\left(x_c(\zeta), y_c(\zeta), z_c(\zeta)\right), \quad \zeta\in[-\pi,\pi]
\end{equation}
with
\begin{equation}\label{eq:centerline}
	\left.
	\begin{aligned}
		x_c(\zeta) &= a\cos (\zeta)+a, \\
		y_c(\zeta) &= h\exp \left(-\frac{\zeta^2}{2}\right), \\
		z_c(\zeta) &= \begin{cases}
			-b\left(\zeta-\frac{\pi}{4}\right)\tan\varphi-\frac{\sqrt{2}}{2}b, & -\pi\leq\zeta<-\frac{\pi}{4}, \\ 
			b \sin \zeta, & -\frac{\pi}{4}\leq \zeta \leq \frac{\pi}{4}, \\  
			b\left(\zeta-\frac{\pi}{4}\right)\tan\varphi+\frac{\sqrt{2}}{2}b, & \frac{\pi}{4}<\zeta\leq\pi,
		\end{cases}
	\end{aligned}
	\right\}
\end{equation}
where $ h$ represents the height of the hairpin vortex, $a = b = h/2 $ are the characteristic lengths in the streamwise and spanwise directions ensuring a rounded transition at the vortex head to approximate the $\Omega$-shaped head described in previous studies on hairpin vortices \citep{Zhou1999Mechanisms}, and $ \varphi = \pi/6 $ is the characteristic inclination angle of the vortex legs \citep{Robinson1991Coherent}. As shown in figure~\ref{fig:hairpin}(a), the parametric equation consists of three segments, with the elliptical vortex head connecting the two vortex legs. The smoothly connected centerline eliminates the sharp angles found in previous studies where attached eddies were composed of only straight-line segments \citep{Marušić1995A,deSilva2016Uniform}. In the $x$-$y$ plane, the inclination angle of the vortex increases exponentially with the distance from the wall-normal direction and maintains an overall tilt of $ 45^\circ $ (see figure~\ref{fig:hairpin}(b)) to conform with experimental observations \citep{Head1981New,Haidari1994The} and recent quantitative analyses based on spectral coherence~\citep{Deshpande2019Streamwise,Cheng2022Streamwise}.

%Gaussain profile
The vorticity field of the hairpin vortex is constructed around the centerline $\mc C$ given by \eqref{eq:centerline} (as shown in figure~\ref{fig:hairpin}(c)), using the curved cylindrical coordinate system $(\zeta, \rho, \theta)$ surrounding the vortex centerline $\mc C$. The vorticity distribution for hairpin vortex tubes with variable thickness~\citep{Xiong2020Effects,Shen2023Role,Shen2024Designing} is specified as
\begin{equation}\label{eq:tubeconstruction}
	\boldsymbol{\omega}(\zeta, \rho, \theta)=\Gamma f(\zeta,\rho)\left[\underbrace{\boldsymbol{e}_{s}}_{\text {flux}}+\underbrace{\frac{\rho}{\sigma(\zeta)(1-\kappa_c(\zeta) \rho \cos \theta)}\frac{\mathrm{d} \sigma(\zeta)/\mathrm{d} \zeta}{\mathrm{d} s(\zeta) /\mathrm{d} \zeta}\boldsymbol{e}_{\rho}}_{\text {tube thickness}}  \right]
\end{equation}
where $\Gamma$ denotes the circulation, $\kappa_c$ the curvature of vortex centerline, $\sigma$ the vortex core size, and the Gaussian kernel function
\begin{eqnarray}\label{eq:Gaussian}
	f(\zeta,\rho)=\left\lbrace \begin{aligned}
		&\frac{1}{2\pi\sigma(\zeta)^2}\exp\left[ \frac{-\rho^2}{2\sigma(\zeta)^2}\right] ,\quad \rho \in [0,R_v),\\
		&0,\quad \rho \in [R_v,+\infty)
	\end{aligned} \right.
\end{eqnarray}
follows the Burgers vortex model. Here, the arc-length parameter is defined as $s(\zeta)=\int_{-\pi}^\zeta\left|\boldsymbol{x}_c^{\prime}(\xi)\right| \mathrm{d} \xi$, $\rho$ represents the radial distance from $\mc C(s)$, and $\theta$ denotes the azimuthal angle from $\bs N(s)$ in the plane $S_C$ spanned by $\bs N(s)$ and $\bs B(s)$. These unit vectors, along with the tangent $\bs T(s)$, form the Frenet--Serret frame on $\mc C$. The two terms in \eqref{eq:tubeconstruction} correspond to the vorticity flux and tube thickness components of $\boldsymbol{\omega}$, respectively.
The vector field constructed by \eqref{eq:tubeconstruction} is proved to be divergence-free~\citep{Shen2023Role,Shen2024Designing}.
The radius $R_v$ of the constructed tubular region is large enough to ensure that almost all circulation is included.

The Lagrangian dynamics of hairpin vortex formation in wall turbulence \citep{Zhao2016Vortex} suggest that vortex tubes near the wall are thicker. Accordingly, we define the vortex core size as  
\begin{equation}\label{eq:sigmaij}
	\sigma(\zeta) = \sigma_0\left[ 1 - C_\sigma\cos(\zeta) \right],
\end{equation}  
where the core size decreases with height above the wall and the core variation coefficient is set to $C_\sigma = 0.3$, which is found to be associated with small-scale and detached structures but largely insensitive to flow statistics (see \S\,\ref{subsec:attdet} and Appendix~\ref{app:sensitivity}). The average core size of the vortex tubes is set to $\sigma_0 = 0.05h$, consistent with the parameters used in the attached-eddy model described in the literature \citep{Perry1995A}. The geometry of a single hairpin vortex is depicted by the vortex surface in figure~\ref{fig:hairpin}(c). Here, vortex surfaces represent the coaxial envelope of vortex lines. To analyze the complex vortex topology, we use the normalized vortex surface field (VSF), $\phi_v = 2\pi\sigma^2 f$, as an effective tool \citep{Yang2010On,Yang2023Applications,Shen2024Designing}.

\subsection{Attached-eddy model}\label{subsec:AEM}

Hairpin vortices of various length scales are distributed along the wall surface according to the attached-eddy hypothesis~\citep{Marusic2019Attached}. In this model, the distance of a hairpin vortex from the wall is proportional to its characteristic length scale, and vortices at different scales follow distinct population densities.
Attached-eddy models describe how these vortices are arranged: they can be randomly distributed~\citep{Perry1995A} or form coherent structures~\citep{Marusic2001On}. The latter is supported by experimental and numerical observations~\citep{Adrian2000Vortex, Zhou1999Mechanisms}, which reveal that hairpin vortices often align spatially in wall turbulence, forming coherent vortex packets along the flow direction.

We determine the scale range, hierarchical levels, and number of hairpin vortex packets in wall turbulence based on the friction Reynolds number, $\Rey_\tau \approx 0.09 \Rey^{0.88}$ \citep{Pope2000Turbulent}, and a boundary layer height of $\delta=1$. The Reynolds number is defined as $\Rey \equiv 2\delta \bar{U} / \nu$, where $\bar{U} \equiv \frac{1}{\delta} \int_0^\delta \langle U \rangle \mathrm{d}y = 1$ represents the bulk velocity, and $\nu$ is the kinematic viscosity. It should be noted that the choices $\delta = 1$ and $\bar{U} = 1$ result from nondimensionalization for analytical convenience, and these quantities may take arbitrary physical values in actual applications.
To ensure consistency with known turbulence structures, these parameters are carefully designed and calibrated using attached-eddy models that have been extensively studied in the literature. By systematically incorporating and refining these models, we construct a self-similar hierarchy of hairpin vortex packets, ensuring that the vortex distribution and scaling behaviour align with established theoretical frameworks.
	
At hierarchical vortex packet level $i$,  
vortex packets exhibit characteristic heights
\begin{equation}\label{eq:hi}
	%h_i = h_{\max } \alpha^{N_v(i-1)}, \quad i=1,2,\ldots,N_p
	h_i = h_{\min } \alpha^{i-1}, \quad i=1,2,\ldots,N_p,
\end{equation}
where $h_{\min} = 200 \Rey_\tau^{-1}$ is the minimum vortex packet height, rather than a single hairpin vortex. The actual minimum height of individual hairpin vortices is taken to be $100 \Rey_\tau^{-1}$, consistent with the smallest self-similar structures reported in \citet{Woodcock2015Statistical}.
The number of hierarchical vortex packet levels, $N_p$, follows a logarithmic-linear relationship~\citep{Woodcock2015Statistical}
\begin{equation}\label{eq:Np}
	%N_p = \operatorname{round}(a_1 \log \Rey_\tau + a_0)
	N_p=1+\left\lfloor\log _2\left(\delta / h_{\min }\right)\right\rfloor.
\end{equation}
Subsequent hierarchical levels display geometric progression in their length scale with the packet scaling factor $\alpha=2$ \citep{perry1982mechanism}.

The population density of packets of scale $i$, defined as the number of packets attached to the wall per unit area, follows a $-2$ power law
\begin{equation}\label{eq:Mi}
	M_i = \kappa h_i^{-2} \propto \alpha^{-2i},
\end{equation}
where $\kappa=0.2$ is a constant. These vortex packets are randomly distributed along the wall while maintaining a minimum spacing of $2h_i$ between any two eddies of the same height to preserve the sub-Gaussian behaviour of the streamwise velocity component~\citep{deSilva2016Influence}.
 
\begin{figure}
	\centering
	\includegraphics[width=0.95\linewidth]{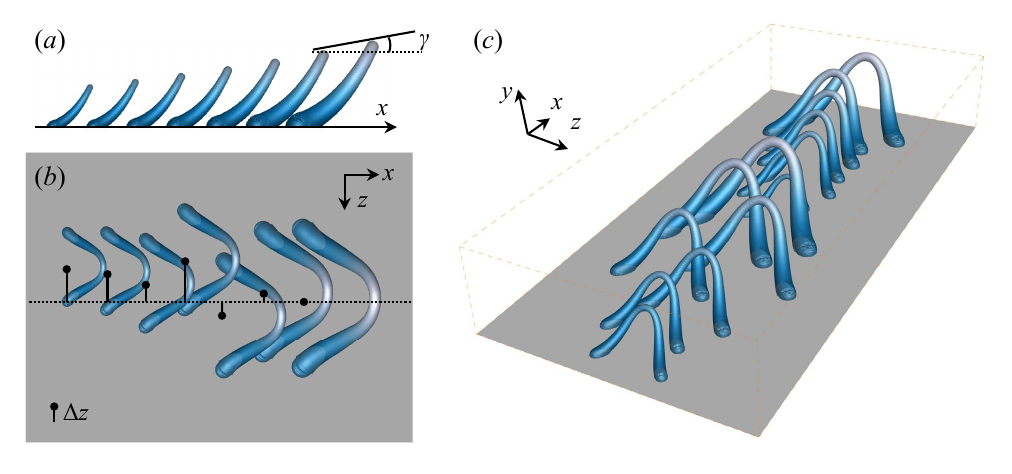}
	\caption{Geometry of vortex packets and wall-coherent superstructures. (a) Alignment of sub-level hairpin vortices along the streamwise direction, exhibiting increasing heights at a growth angle of $\gamma$. (b) Top view of the vortex packet, illustrating the spanwise meandering features. (c) Vortex surface visualization $\phi_v=0.1$ of a typical wall-coherent superstructure.}
	\label{fig:packet}
\end{figure}

Each vortex packet consists of $N_v = 7$ sub-level hairpin vortices arranged along the flow direction~\citep{deSilva2016Uniform}. As illustrated in figure~\ref{fig:packet}(a), these sub-level vortices are evenly distributed along the streamwise length of the packet, defined as $L_i=3h_i$, consistent with earlier AEM-based attempts \citep{Deshpande2021Data} and experimental observations of packet lengths \citep{Adrian2007Hairpin}. We adopt $N_v=7$ as a baseline because it provides an inter-head spacing $L_i/(N_v-1)= 0.5 h_i$, yielding a compact yet representative packet. In practice, $N_v$ can be treated as a tunable parameter or vary with scale \citep{Adrian2000Vortex}. The heights of individual sub-level vortices within a level-$i$ packet follow  
\begin{equation}\label{eq:hij}
	h^{(j)}_i = h_i \alpha^{(1-j)/(N_v-1)}, \quad j=1,2,\ldots,N_v,
\end{equation}  
ensuring a smooth transition of hairpin vortex heights across different vortex packet levels. In this configuration, the hairpin vortices collectively form an inclined hairpin ramp from upstream to downstream, with a growth angle reaching up to $\gamma \approx 12^\circ$, consistent with experimental observations~\citep{Adrian2000Vortex}. The circulation of each hairpin vortex is given by  
\begin{equation}\label{eq:Gammaij}
	\Gamma^{(j)}_i = C_\Gamma u_\tau h_i,
\end{equation}  
where $C_\Gamma=2$ is the circulation coefficient and $u_\tau = \nu \Rey_\tau / \delta$ represents the friction velocity.

A key advantage of constructing wall turbulence based on vortex building blocks is the flexibility to phenomenologically adjust the spatial arrangement of hairpin vortices to incorporate theoretical results and experimental/numerical observations. Considering the meandering features of large-scale structures in turbulent boundary layers~\citep{Hutchins2007Evidence}, we introduce the spanwise meandering of hairpin vortex packets (see figure~\ref{fig:packet}(b)).
Each hairpin vortex within a level-$i$ packet undergoes a random spanwise displacement
\begin{equation}\label{eq:zi}
	\Delta z_i = \sigma_m \xi h_i, \quad \xi \sim \mathscr{U}(-1, 1),
\end{equation}
where $\sigma_m=0.5$ is a constant, and $\xi$ is a uniform random variable in the range $(-1,1)$, in agreement with experimental~\citep{Hutchins2007Evidence} and numerical~\citep{Adrian2002Observation,Hwang2022Meandering} observations.

In addition to these self-similar vortex packets, we incorporate wall-coherent superstructures within the largest vortex packets, also referred to as VLSMs~\citep{Kim1999Very,Chandran2020Spectral}, to supplement the dynamics of the outer layer.
Building on the concept that vortex packets align in the streamwise direction to form extended structures, we model these superstructures by combining two of the largest packets. 
This approach follows earlier extensions of AEM, in which VLSMs are constructed by aggregating vortex packets to capture their streamwise coherence and scale separation from near-wall motions~\citep{Chandran2020Spectral,Deshpande2021Data}.
Empirical support for this modeling rationale is provided by \citet{Deshpande2023Evidence}, who demonstrate that VLSMs comprise smaller, geometrically self-similar coherent motions.

The height of the modelled superstructure is set to match the boundary layer thickness, $h_{SS} = \delta$, ensuring consistency with the overall flow organization, as illustrated in figure~\ref{fig:packet}(c).
Furthermore, the circulation of each hairpin vortex within VLSMs is prescribed as $ \Gamma_{SS} = 0.1 \bar{U} h_{SS} $, where the velocity scale is chosen as the bulk velocity $ \bar{U} $ rather than the friction velocity $u_\tau$ used for self-similar vortex packets. These superstructures span the entire boundary layer height and dominate the outer layer, highlighting the influence of large-scale outer-layer structures over near-wall turbulence characteristics.

\subsection{Construction method of flow fields}\label{subsec:construction}

To satisfy the no-penetration condition at the wall, we construct hairpin vortices directly on the wall, and introduce symmetric virtual vortices in the mirror space relative to the wall. The influence of the wall is modelled through the induced velocity of these virtual vortices within the computational domain. The vorticity field is assembled by summing contributions from all hairpin vortices, following \eqref{eq:tubeconstruction}, using a numerical algorithm that transforms Cartesian coordinates $\boldsymbol{x}$ into $(s, \rho, \theta)$. This approach has proven highly robust, producing a smooth vorticity field even in cases of complex vortex tube interactions, including self-intersections~\citep{Shen2024Designing}. The fluctuating velocity field is then computed from the vorticity via the Biot--Savart law in Fourier space. To better resolve the wall flow, we interpolate the data onto a non-uniform grid based on Chebyshev polynomials~\citep{Wang2024Transition}. 

In our approach, the induced velocity field of the constructed eddies $\tilde{U}$, along with an extra uniform bulk velocity $\bar{U}=1$, directly determine the wall-normal distribution of the turbulent velocity profile
\begin{equation}
	U = (\tilde{U} + \bar{U}) (1 - \exp\left(-y / A\right) ).
	\label{eq:damping}
\end{equation}  
Note that a van Driest-type damping function is introduced to enforce the no-slip boundary condition, where the constant 
\begin{equation}
A=\frac{\delta^2\langle \tilde{U}\rangle_{y=0}}{\Rey_\tau^2 \nu}
\end{equation} 
depends on $\Rey_\tau$, ensuring accurate reproduction of the wall shear stress. 

\begin{figure}
	\centering
	\includegraphics[width=\linewidth]{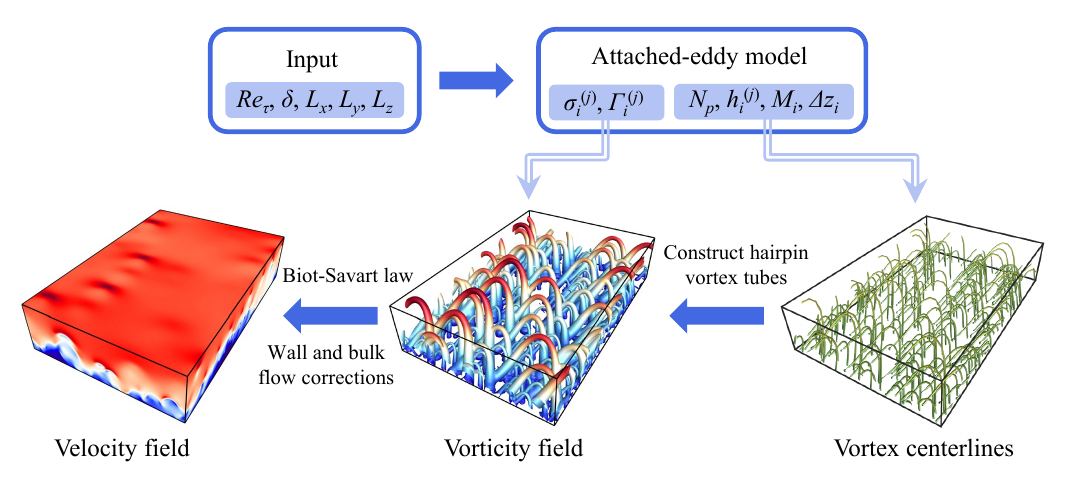}
	\caption{Construction of synthetic wall-attached turbulence. The input parameters include the prescribed friction Reynolds number $\Rey_\tau$, boundary layer thickness $\delta$, and the dimensions of the computational domain $L_x \times L_y \times L_z$. Based on the attached-eddy model, key structural properties are determined, including the hierarchical level number of vortex packets $N_p$ in \eqref{eq:Np}, the height of individual hairpin vortices $h_i^{(j)}$ in \eqref{eq:hi}, population density $M_i$ in \eqref{eq:Mi}, and spanwise meandering $\Delta z_i$ in \eqref{eq:zi}. These parameters define the centerlines of the hairpin vortices. Next, the vorticity field of the hairpin vortices is constructed based on their centerlines, circulation strengths $\Gamma_i^{(j)}$ in \eqref{eq:Gammaij}, and core size distribution $\sigma_i^{(j)}$ in \eqref{eq:sigmaij}. Finally, the velocity field of the synthetic wall turbulence is obtained by applying the Biot--Savart law along with wall and bulk flow corrections in \eqref{eq:damping}.}
	\label{fig:flowchart}
\end{figure}

Figure~\ref{fig:flowchart} provides an overview of the construction process for SWAT, summarizing the key steps involved. This process is based on the attached-eddy model, where the structural properties of vortex packets are determined by input parameters such as the friction Reynolds number and boundary layer thickness. These properties include the hierarchical distribution of hairpin vortices, their population density, and spanwise meandering. The vorticity field is then reconstructed using the centerlines, circulation strengths, and core size distributions. Finally, the velocity field is computed using the Biot–Savart law, with corrections for wall and bulk flow. In our approach, the only required input parameter for a given channel flow within a specified computational domain is the friction Reynolds number $\Rey_\tau$. All other flow characteristics, including the hierarchical vortex structures, their spatial distributions, and the velocity field, are systematically derived from this single parameter.

Compared to DNS, the computational cost of the SWAT method is almost negligible. Table~\ref{tab:cost} summarizes key metrics for all SWAT cases presented in this paper, covering friction Reynolds numbers ranging from 1,000 to 10,000. These metrics include the number of computational grid points, the number of vortices, and the CPU hours required for each calculation. The smallest hairpin vortex height is resolved with at least five grid points in the wall-normal direction, ensuring an accurate representation of coherent structures. The computational expense of SWAT primarily depends on the number of grid points and the total number of self-similar vortices, and thus increases with the specified Reynolds number. Nevertheless, the method’s efficient numerical construction keeps the overall computational cost significantly lower than DNS.

\begin{table*}
\caption{Summary of computational parameters and cost for synthetic wall-attached turbulence in half channel flow at various friction Reynolds numbers $\Rey_\tau$. The computational domain is fixed at $8\pi \times 1 \times 3\pi$ in the streamwise ($x$), wall-normal ($y$), and spanwise ($z$) directions, respectively. The table lists the number of grid points in each direction ($N_x$, $N_y$, $N_z$), the total number of self-similar vortices synthesized, and the CPU hours required for constructing each flow field.}
\setlength{\tabcolsep}{2.5mm}
\begin{ruledtabular}
\begin{tabular}{cccccc}
$\Rey_\tau$	& $ N_x$  &  $ N_y$ & $N_z$ & Number of vortices & CPU hours \\ \hline
1,000 & 1024 & 64 & 384 & 11536 & 0.3 \\
2,000 & 1024 & 128 & 384 & 44695 & 1.1 \\
5,200 & 2048 & 256 & 768 & 299250 & 7.8 \\
10,000 & 2048 & 512 & 768 & 1105762 & 13.9 \\
\end{tabular}
\end{ruledtabular}
\label{tab:cost}
\end{table*}

\section{Statistics of synthetic wall-attached turbulence}\label{sec:results}

\subsection{Mean velocity and Reynolds stresses}\label{subsec:LS}
Taking turbulent channel flow as a representative case, we leverage DNS data~\citep{Lee2015Direct,Oberlack2022Turbulence} to systematically assess the statistical properties and structural characteristics of the constructed SWAT. 
We first generate a synthetic wall-bounded flow within a computational domain of $ L_x \times L_z = 8\pi \times 3\pi$,  with a prescribed channel half-height of $\delta = 1$, a Reynolds number of $\Rey_\tau = 1000$. This domain size is selected to be consistent with the reference DNS database, allowing direct comparison of statistical profiles and ensuring sufficient space to accommodate the largest coherent structures.

\begin{figure}
	\centering
	\includegraphics[width=\linewidth]{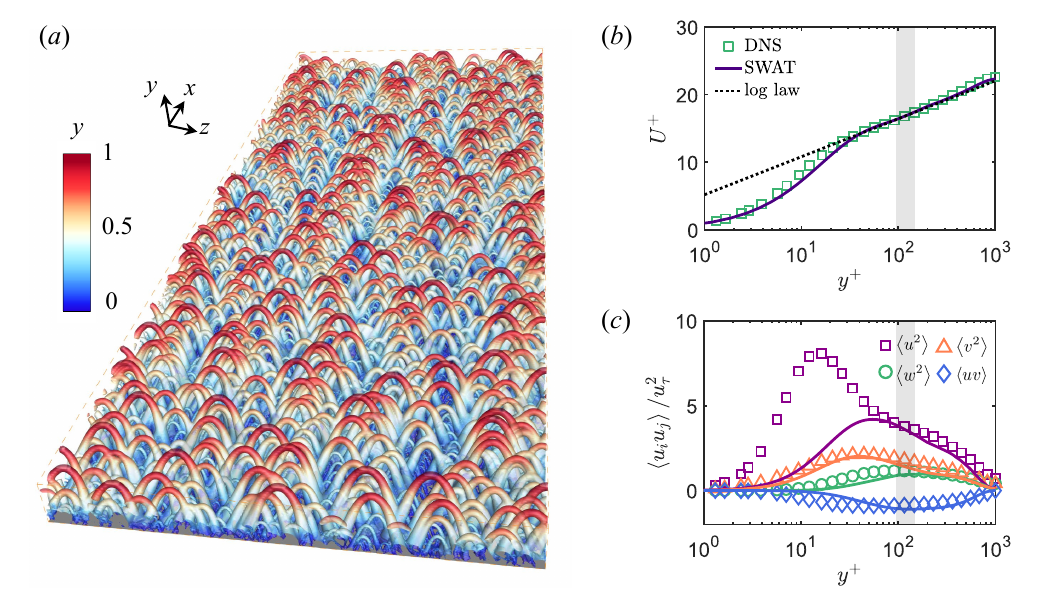}
	\caption{Structure and statistics of the synthetic wall-attached turbulence (SWAT) for $\Rey_\tau=1000$. (a) Visualization of vortex surfaces in SWAT, displaying hierarchically attached vortex packets with spanwise meandering features. The vortex surfaces are colour-coded according to the wall distance. (b,c) Comparison of (b) mean velocity and (c) Reynolds stress profiles between SWAT and DNS data. Symbols represent DNS data from \citet{Lee2015Direct}, while solid lines in matching colours represent SWAT. The gray area marks a strictly logarithmic region~\citep{Marusic2013On}, defined as $3\Rey_\tau^{1/2}<y^+<0.15\Rey_\tau$.} 
	\label{fig:SWAT}
\end{figure}

To visualize the generated turbulence, we employ the VSF of $\phi_v = 0.1$, as illustrated in figure~\ref{fig:SWAT}(a). The synthetic flow field, constructed using our numerical methods and carefully designed vortex elements, exhibits structural features that closely resemble those observed in both experimental measurements and high-fidelity DNS data. Notably, vortices spanning a wide range of length scales remain attached to the wall, forming a complex multiscale structure of a vortex forest. This hierarchical organization of vortices is a key characteristic of wall turbulence and reflects important structural features observed in real turbulent boundary layers, although the present simulations do not capture the full dynamical interactions among these scales.

Our primary objective is to determine whether such vortex assemblies can faithfully reproduce the mean velocity profile of wall turbulence, particularly the scaling law in the logarithmic region, which is fundamental to the physics of wall turbulence. Figure~\ref{fig:SWAT}(b) presents the mean velocity profile of SWAT, obtained through ensemble-averaging. Compared to DNS data, the SWAT model accurately captures the mean velocity distribution across different wall distances in turbulent flow. It is important to emphasize that no pre-averaged flow data from numerical simulations or experiments were used in the construction of SWAT.  

In the near-wall region (with the wall unit $ y^+ < 5 $), the velocity profile is primarily determined by the imposed damping function in \eqref{eq:damping}, which controls the flow in the viscous sublayer. In the logarithmic region ($y^+ > 30$), the emergence of the log law in the mean velocity profile arises entirely from the collective contribution of all the hairpin vortices, without any explicit enforcement. The buffer layer ($ 5 < y^+ < 30 $) provides a smooth transition between the viscous sublayer and the logarithmic region, further validating the self-consistent nature of the synthetic turbulence model.

Reynolds stresses characterize the momentum transfer induced by turbulent eddies. Figure~\ref{fig:SWAT}(c) compares the Reynolds stress components obtained from SWAT with those from DNS of channel flow. The SWAT effectively reproduces the Reynolds stress distributions across different wall-normal locations, demonstrating the expected anisotropic characteristics of wall turbulence. The components $ \langle v^2 \rangle $, $ \langle w^2 \rangle $, and $ \langle uv \rangle $ exhibit strong agreement with DNS data, indicating that the vortex structures incorporated in SWAT capture key features of momentum transfer.  
The gray region in figures~\ref{fig:SWAT}(b,c) highlights the strictly logarithmic region, defined as $ 3\Rey_\tau^{1/2} < y^+ < 0.15\Rey_\tau $~\citep{Marusic2013On}, based on higher-order statistical considerations. Within this region, the streamwise component $\langle u^2 \rangle$ captures the expected scaling behaviour but underpredicts the near-wall peak. This underestimation is likely due to the partial loss of small-scale energy associated with the absence of viscosity-dominated eddies, highlighting the need for improved representation of near-wall dynamics.

\subsection{Energy spectrum and higher-order statistics}\label{subsec:HS}

\begin{figure}
	\centering
	\includegraphics[width=\linewidth]{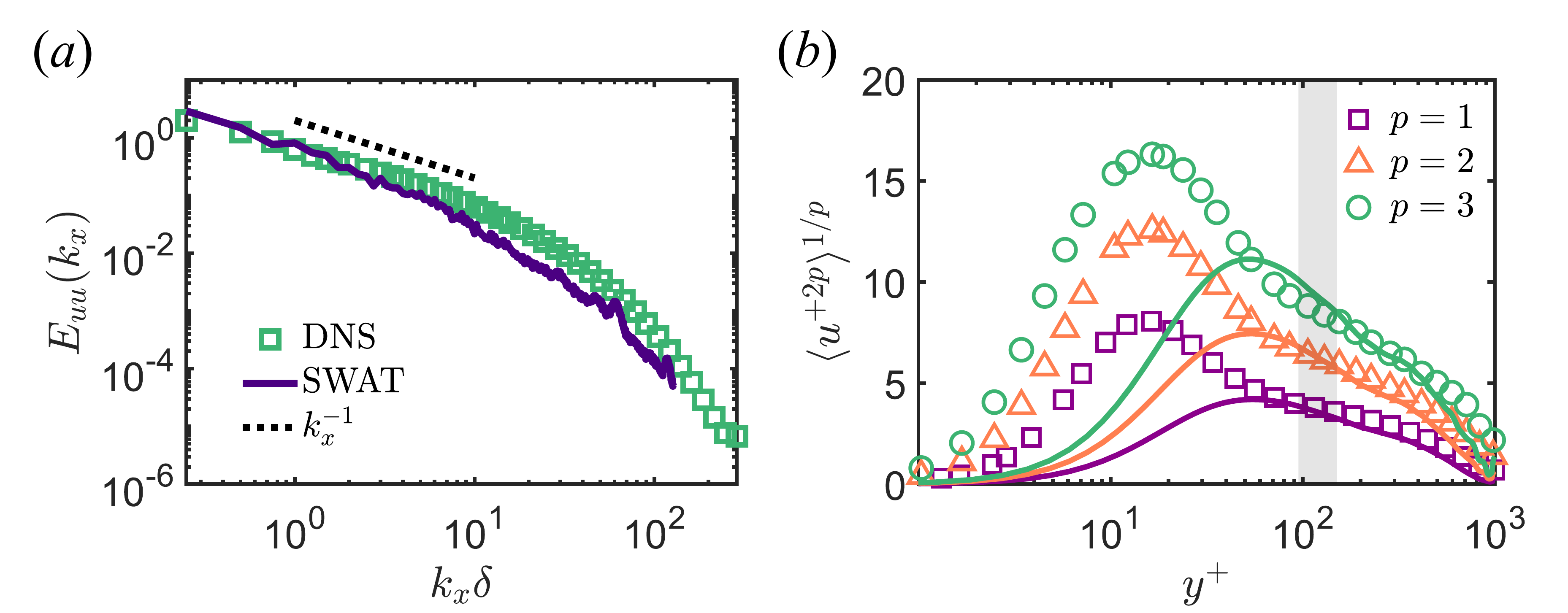}
	\caption{Streamwise energy spectra and higher-order statistics of SWAT and DNS at $\Rey_\tau=1000$. (a) Streamwise energy spectra at $y^+=3.9Re_\tau^{1/2}$. (b) Profiles of higher-order statistics for the streamwise velocity $ \langle [(u-\langle u \rangle)/u_\tau]^{2p}\rangle^{1/p} $. Symbols represent DNS data from \citet{Lee2015Direct}, while solid lines in corresponding colors denote SWAT results.}
	\label{fig:highorder}
\end{figure}

The fidelity of the SWAT model is first validated through spectral analysis. As shown in figure~\ref{fig:highorder}(a), the streamwise energy spectrum of the synthetic turbulence in the logarithmic region exhibits the streamwise wavenumber $k_x^{-1}$ scaling in the low-wavenumber (large-scale) region, consistent with the energy distribution observed in canonical wall turbulence\cite{Nickels2005Evidence,Baars2019Data1}. This agreement highlights SWAT’s capability to resolve the dominant energy-containing motions that govern turbulent transport processes in the logarithmic region. The premultiplied form likewise shows the emergence of an energetic plateau, with its Reynolds-number dependence examined in the next subsection.

The hairpin-vortex-based SWAT model also captures the higher-order dynamics characteristic of wall-bounded turbulence in the logarithmic region.
Conventional synthetic turbulence models often rely on ad hoc parameter tuning to match first- or second-order statistics, yet they frequently overlook the higher-order statistical complexity. The SWAT framework overcomes this limitation by embedding physically meaningful vortical structures that naturally encode multiscale interactions and nonlinear dynamics. As shown in figure~\ref{fig:highorder}(b), SWAT accurately reproduces the logarithmic scaling of even-order statistical moments of streamwise velocity fluctuations, $ \langle [(u-\langle u \rangle)/u_\tau]^{2p}\rangle^{1/p} $, with respect to the distance from the wall within the logarithmic sublayer~\citep{Meneveau2013Generalized}. It is worth noting that the higher-order statistics exhibit poor agreement in the near-wall region, likely due to the absence of viscosity-dominated eddies. This corresponds to the underestimation of small-scale features observed in figure~\ref{fig:highorder}(a).

\subsection{Towards higher Reynolds numbers}\label{subsec:Re}

The nature of coherent structures in high-Reynolds-number wall turbulence remains a topic of active debate~\citep{Wu2009Direct, Smits2011High, EitelAmor2015Hairpin}. A central question is whether simple hairpin vortices can still provide an adequate modeling framework at extremely high Reynolds numbers. This poses a fundamental challenge for our model, as the increasing complexity of turbulence may demand a more sophisticated representation of vortex interactions and energy transfer mechanisms.  

\begin{figure}
	\centering
	\includegraphics[width=0.9\linewidth]{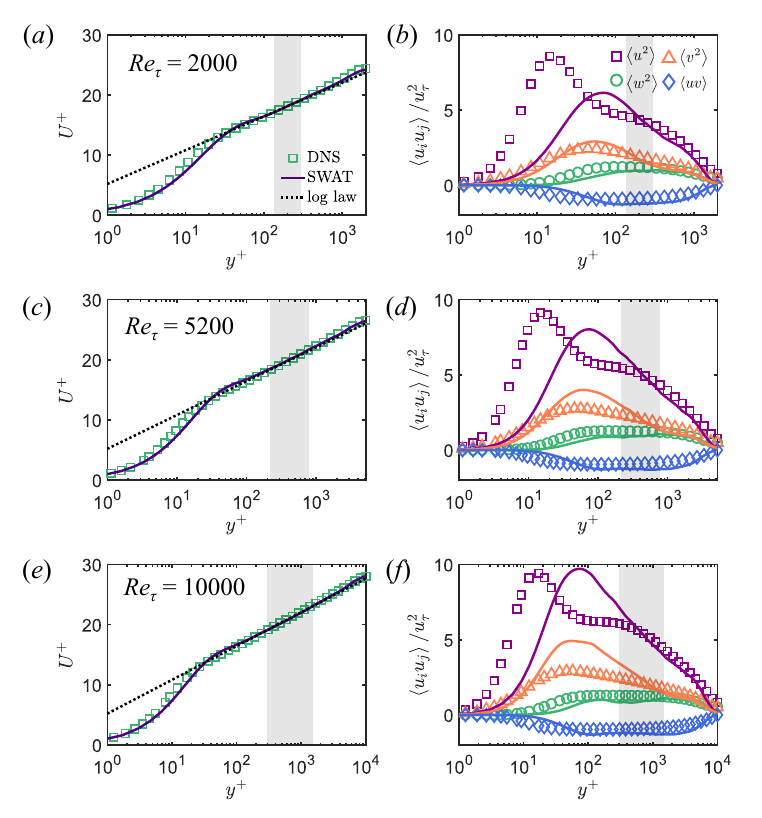}
	\caption{Comparison of (a,c,e) mean velocity profiles and (b,d,f) Reynolds stress profiles between synthetic wall-attached turbulence (SWAT) and DNS data for Reynolds numbers (a,b) $\Rey_\tau = 2000$, (c,d) $5200$, and (e,f) $10000$. Symbols represent DNS data from \citet{Lee2015Direct} ($\Rey_\tau = 2000, 5200$ ) and \citet{Oberlack2022Turbulence} ($\Rey_\tau = 10000$ ), and solid lines in matching colours indicate SWAT results. The shaded grey region highlights the log-law region $3\Rey_\tau^{1/2}<y^+<0.15\Rey_\tau$.}
	\label{fig:statistics}
\end{figure}

To investigate this, we generate synthetic wall turbulence at progressively higher Reynolds numbers, specifically at $Re_\tau = 2000 $, $ 5200 $, and $ 10000 $. Across this range, SWAT consistently demonstrates its ability to accurately reproduce the mean velocity profile throughout the near-wall, buffer, and logarithmic regions in figures~\ref{fig:statistics}(a,c,e). This agreement underscores the model's ability to capture the key statistical features of turbulent boundary layers, even at very high Reynolds numbers.

Regarding the Reynolds stress distribution, all second-order moments exhibit the expected logarithmic behaviour in the log-law region across all tested Reynolds numbers in figures~\ref{fig:statistics}(b,d,f). Compared to DNS data, SWAT effectively captures the $ \langle w^2 \rangle $ and $ \langle uv \rangle $ components, which are crucial for characterizing turbulence anisotropy and momentum transfer. Additionally, the model successfully replicates the logarithmic growth trends of the peak values of $ \langle u^2 \rangle $ and $ \langle v^2 \rangle $ with increasing Reynolds number \citep{LozanoDuran2014Effect,Lee2015Direct}, although the growth rate is slightly overestimated. 

\begin{figure}
	\centering
	\includegraphics[width=0.95\linewidth]{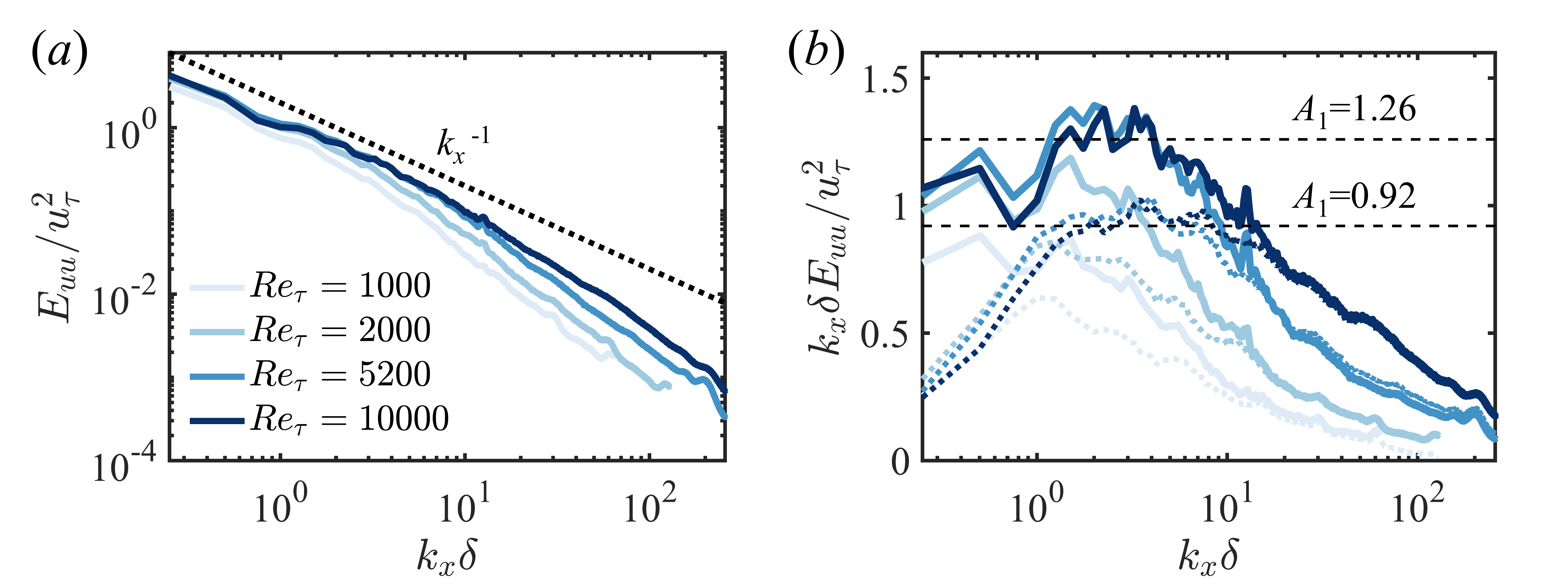}
	\caption{(a) Streamwise energy spectra in SWAT at $\Rey_\tau = 1000, 2000, 5200,$ and $10000$ at the logarithmic region $y^+ = 3.9\Rey_\tau^{1/2}$. (b) Corresponding premultiplied spectra. The dotted lines of matching colors represent the self-similar packet contributions with superstructures removed, at the respective Reynolds numbers. Each spectrum is ensemble-averaged over 250 SWAT flow fields.}
	\label{fig:premultiplied}
\end{figure}

We further compare the streamwise energy spectra in the logarithmic region at different Reynolds numbers in figure~\ref{fig:premultiplied}(a), all of which exhibit the characteristic $k_x^{-1}$ scaling. The corresponding premultiplied spectra are shown in figure~\ref{fig:premultiplied}(b), where the range of the $k_x^{-1}$ region is seen to extend with increasing Reynolds number, indicating a progressively wider self-similar scaling regime. The dotted lines in figure~\ref{fig:premultiplied}(b) denote the streamwise spectra of SWAT with the superstructures removed, highlighting the self-similar contributions in the flow. These spectra exhibit excellent Reynolds-number convergence, and the extent of the $k_x^{-1}$ region likewise grows with Reynolds number. The spectra plateau at the Townsend–Perry constant $A_1$, whose magnitude provides a key test of the AEH \citep{Baars2019Data2}. Different studies have reported values of $A_1=0.92$ \citep{Nickels2005Evidence} and $A_1=1.26$ \citep{Hultmark2012Turbulent, Marusic2013On, Orlu2017Reynolds}, which are indicated in figure~\ref{fig:premultiplied}(b). Our results show good agreement with these estimates.

These results indicate that SWAT remains a robust framework for modeling the statistical properties of wall turbulence across a broad range of Reynolds numbers.

\section{Insights into wall turbulence structure from SWAT}

Owing to its incorporation of highly tunable structural elements, the SWAT approach serves as a versatile testbed for investigating instantaneous coherent structures from the perspective of building blocks. It enables controlled manipulation of vortex geometry and organization, thereby facilitating both the validation of observational findings and conceptual models, and offers mechanistic insights into the coherent organization of wall turbulence.

\subsection{Hairpin vortices and attached/detached structures}\label{subsec:attdet}

A central step in SWAT is the construction of realistic hairpin vortices that act as building blocks of wall turbulence. Although these hairpins are defined in terms of vortex surfaces and are therefore wall-attached by construction, their specific configuration enables them to simultaneously represent both attached and detached structures.

\begin{figure}
	\centering
	\includegraphics[width=0.8\linewidth]{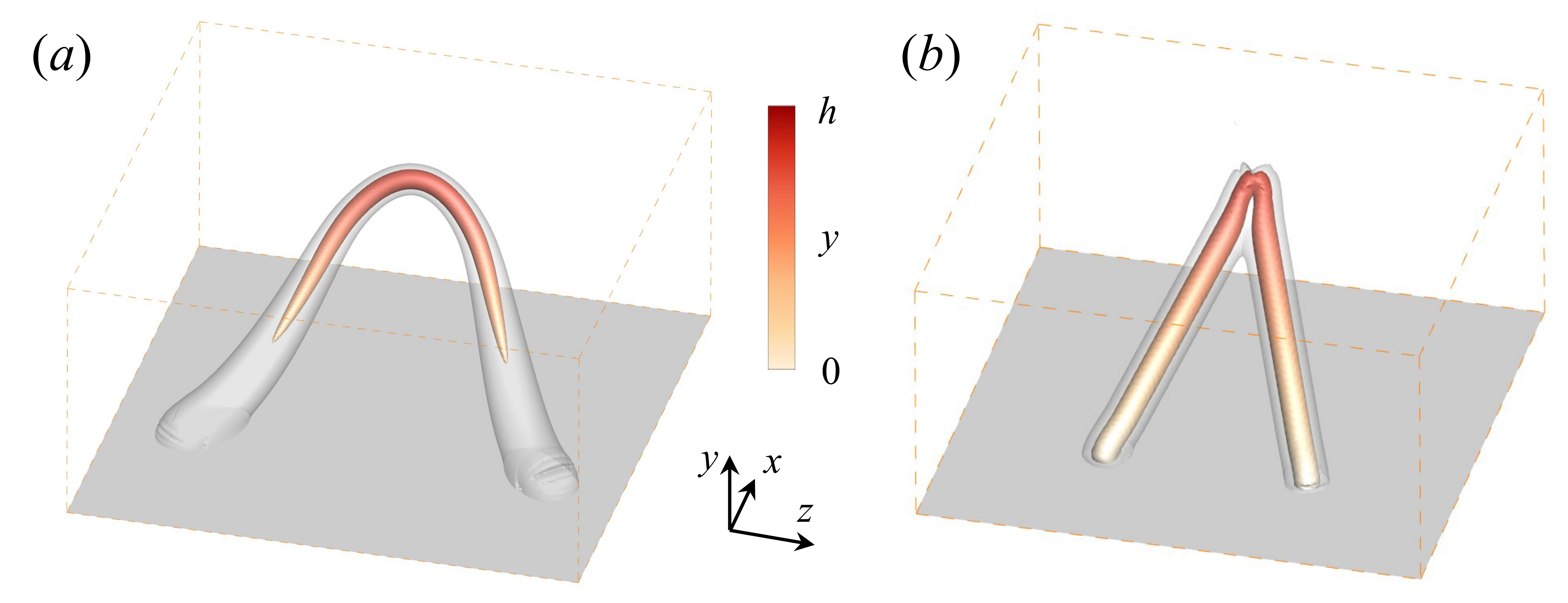}
	\caption{Visualization of the vortex surface and vorticity magnitude of a single hairpin vortex: (a) the present improved hairpin vortex model with an $\Omega$-shaped head and variable core size, and (b) the conventional $\Lambda$-shaped hairpin vortex. The colored structure (detached from the wall in (a)) represents the isosurface of vorticity magnitude at $|\boldsymbol{\omega}| = 0.4|\boldsymbol{\omega}|_{\text{max}}$, while the transparent structure attached to the wall corresponds to the vortex surface at $\phi_v = 0.1$.}
	\label{fig:detachedhairpin}
\end{figure}

To formulate such vortices in a physically consistent manner, SWAT adopts a vortex-surface-based representation, differing from conventional AEM-based flow-field constructions that rely directly on velocity or vorticity fields. The hairpin vortex serves as the elementary unit and is characterized by the normalized VSF
\begin{equation}
	\phi_v(\zeta,\rho) = 2\pi \sigma(\zeta)^2 f(\zeta,\rho) \in [0,1].
\end{equation}
Here, the vortex surface is defined as the envelope of vortex lines, with all vortex lines embedded in iso-surfaces of $\phi_v$ \citep{Yang2010On,Yang2011Evolution}. Motivated by observations of coherent structures in real turbulence, we introduce a height-dependent vortex-core thickness $\sigma(\zeta)$ (see \eqref{eq:sigmaij}), such that vortices are thicker closer to the wall. This design, referred to as the improved hairpin vortex model with an $\Omega$-shaped head and variable core size, preserves the overall circulation while concentrating vorticity near the head. Consequently, iso-surfaces of vorticity magnitude deviate from the corresponding vortex surfaces \citep{Zhao2016Evolution,Zhao2016Vortex}. As illustrated in figure~\ref{fig:detachedhairpin}(a), the vortex surface of a hairpin with variable core size remains attached to the wall, whereas the vorticity-magnitude iso-surface detaches from it. In contrast, the conventional $\Lambda$-shaped hairpin vortex without variable core size (figure~\ref{fig:detachedhairpin}(b)) maintains its structure and remains wall-attached under various vortex-identification criteria.

\begin{figure}
	\centering
	\includegraphics[width=\linewidth]{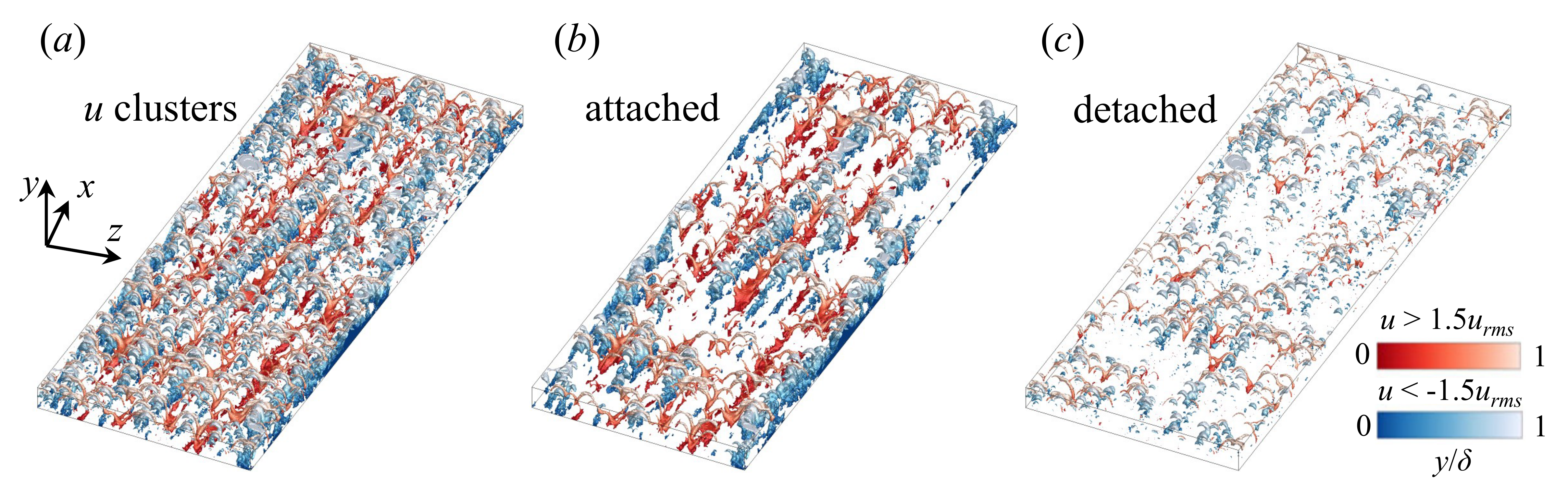}
	\caption{Iso-surfaces of (a) all $u$ clusters, (b) attached $u$ clusters, and (c) detached $u$ clusters in the SWAT at $\Rey_\tau=1000$ with $C_\sigma=0.3$ . Red and blue indicate regions of intense positive and negative streamwise velocity fluctuations, respectively. The iso-surfaces are colour-coded according to the wall distance.}
	\label{fig:ucluster}
\end{figure}

These differences in the underlying vorticity distribution have direct implications for the resulting velocity field.
Although all hairpin vortices are wall-attached when viewed from the perspective of vortex surfaces, the non-uniform vorticity distribution within each configuration means that not all $u$ clusters remain attached. Both attached and detached structures are present. To identify $u$ clusters in SWAT, the connectivity of $u$ is defined using the six orthogonal nearest neighbours of each grid point \citep{delAlamo2006Self,Lozano2012The,Hwang2018Wall,Yoon2019Wall}. Figure~\ref{fig:ucluster}(a–c) shows the structures of all $u$ clusters, attached $u$ clusters, and detached $u$ clusters in SWAT at $\Rey_\tau=1000$ with the core variation coefficient $C_\sigma=0.3$. Intense positive and negative streamwise velocity fluctuations are indicated by red and blue, respectively, and are further coloured by wall-normal distance to distinguish attachment. It is evident that not all $u$ clusters are wall-attached. In addition to the attached structures, SWAT also contains a number of intense detached structures. The attached clusters typically appear as elongated streaks aligned in the streamwise direction, alternating between positive and negative fluctuations, whereas the detached clusters are smaller and more irregular. These observations are consistent with those from boundary-layer DNS \citep{Yoon2019Wall}, demonstrating that even in synthetic wall turbulence constructed from hairpin vortices, appropriately designed hairpin geometries can reproduce both attached and detached structures.

\begin{figure}
	\centering
	\includegraphics[width=0.5\linewidth]{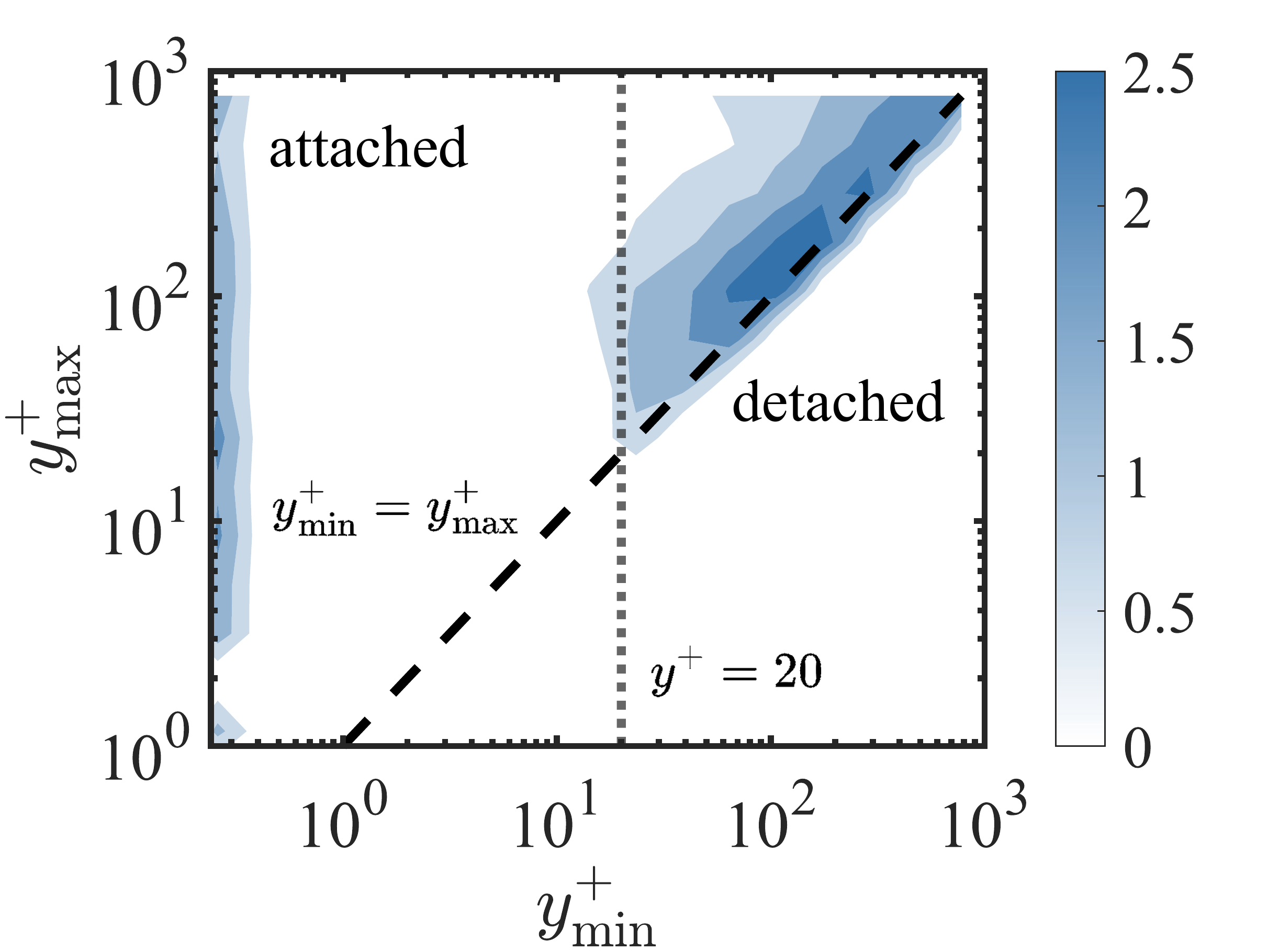}
	\caption{The number and distribution of $u$ clusters per unit wall-parallel area ($L_x \times L_z$) with respect to $y_{min}$ and $y_{max}$ for SWAT at $\Rey_\tau=1000$.}
	\label{fig:ncluster}
\end{figure}

To distinguish between attached and detached structures, the minimum and maximum wall-normal distances of each $u$ cluster ($y_{\min}$ and $y_{\max}$) are measured and analyzed. Figure~\ref{fig:ncluster} shows the number and distribution of $u$ clusters per unit wall-parallel area ($L_x \times L_z$) \citep{Hwang2018Wall} as functions of $y_{\min}$ and $y_{\max}$ at $\Rey_\tau=1000$. Two distinct regimes emerge in the distribution, indicating a classification of clusters into two types: attached structures with $y_{\min}^+ \approx 0$ and detached structures with $y_{\min}^+ > 0$. For attached structures, the minimum wall-normal distance corresponds to the grid point closest to the wall in SWAT, whereas detached structures typically appear at $y^+ > 20$, which is consistent with the existing identification and definitions \citep{delAlamo2006Self,Lozano2012The,Hwang2018Wall,Yoon2019Wall}. 

\begin{figure}
	\centering
	\includegraphics[width=\linewidth]{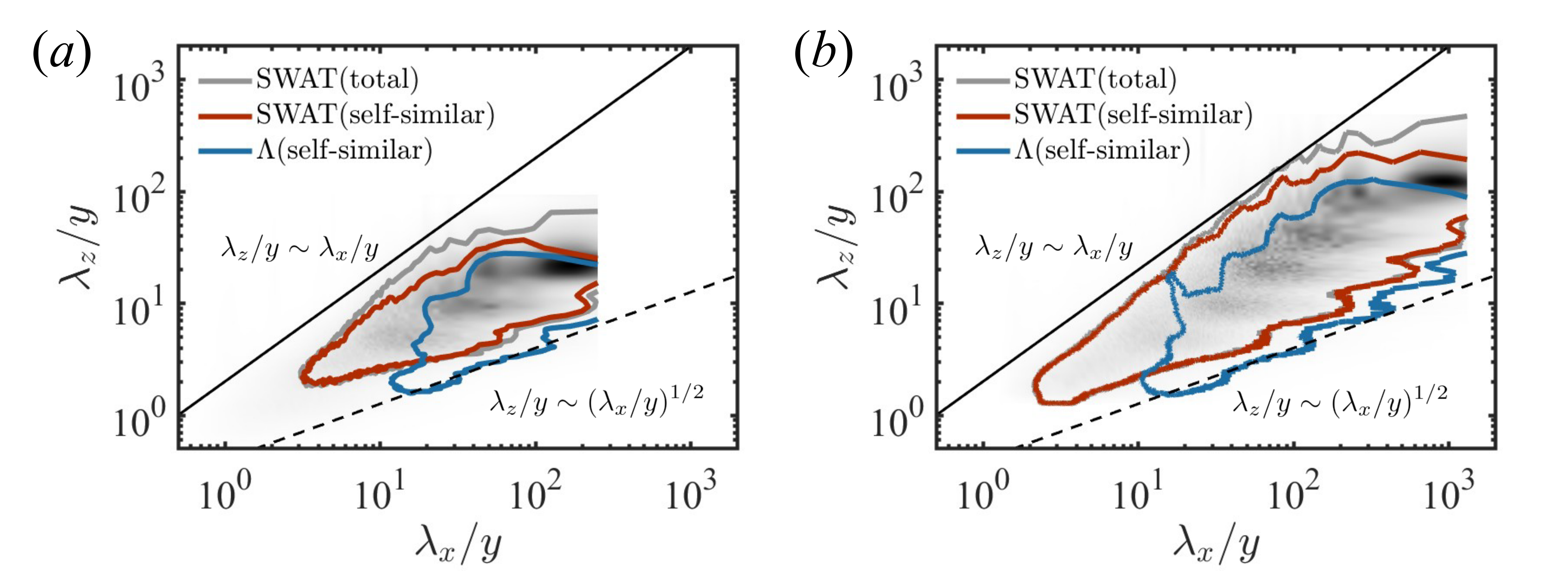}
	\caption{Two-dimensional spectra of the streamwise velocity at $y^+=100$ for (a) $\Rey_\tau=1000$ and (b) $\Rey_\tau=5200$. The spectra are computed from synthetic wall-turbulence fields constructed using the two hairpin configurations shown in figure \ref{fig:detachedhairpin}(a,b), corresponding to the improved $\Omega$-shaped and conventional $\Lambda$-shaped vortices, respectively. The contour level corresponds to $k_x k_y E_{uu}/u_\tau^2 = 0.12$. The black solid and dashed lines denote the $ \lambda_z/y \sim \lambda_x/y$ and $\lambda_z/y \sim ( \lambda_x/y )^{1/2}$ relationships respectively. Blue lines denote the $\Lambda$-shaped self-similar scheme, while red and grey lines represent the current hairpin vortex model including only self-similar eddies (self-similar) and both self-similar eddies and superstructures (total), respectively.}
	\label{fig:2dspectra}
\end{figure}

The modified geometry of the hairpin vortex has an impact on the energy distribution.
We first construct synthetic wall-turbulence fields composed of self-similar vortices using two hairpin configurations, namely the figure \ref{fig:detachedhairpin}(a) improved $\Omega$-shaped vortex with a height-dependent core size and the figure \ref{fig:detachedhairpin}(b) conventional $\Lambda$-shaped vortex. From these two synthetic fields, we compute the two-dimensional energy spectra of the streamwise velocity in the logarithmic region.
The resulting spectra are compared in figure \ref{fig:2dspectra}(a,b) for friction Reynolds numbers $\Rey_\tau=1000$ and $5200$, respectively. The comparison focuses on the self-similar part of the constructed velocity field for both configurations to isolate the effect of the vortex geometry, while the total SWAT field including large-scale superstructures is also shown for reference.
For both configurations, the linear scaling region in the two-dimensional spectra expands with increasing Reynolds number, consistent with the self-similar vortex behaviour discussed by \citet{Chandran2017Two}. However, the optimized $\Omega$-shaped model extends the spectral scaling to smaller scales than the conventional $\Lambda$-shaped scheme, enabling a more accurate reproduction of the small-scale region and the square-root relationship that arises from the height-dependent vortex-core size in the improved model.

In contrast, earlier AEM-based flow fields constructed from conventional $\Lambda$-shaped vortices had to rely on artificially introduced detached vortices to compensate for missing small-scale motions~\citep{Chandran2020Spectral}. These detached vortices were essentially lifted versions of the same $\Lambda$ structures and therefore lacked a physical basis. Our results show that introducing a height-dependent variation of the hairpin vortex core size, consistent with physical observations in real wall turbulence \citep{Zhao2016Vortex}, naturally achieves the required small-scale replenishment without the need for ad hoc vortex additions, thereby providing a self-consistent mechanism rooted in observed wall-turbulence dynamics. 
This mechanism may also account for the detached small-scale motions described as Type-C eddies~\citep{Marušić1995A,Marusic2019Attached}, which likely represent remnants or lifted extensions of attached eddies. The present results suggest that such remnant or lift-off behaviour can arise naturally from the height-dependent variation of vortex-core thickness, providing a continuous link between the attached (Type-A) and detached (Type-C) populations.

\begin{figure}
	\centering
	\includegraphics[width=\linewidth]{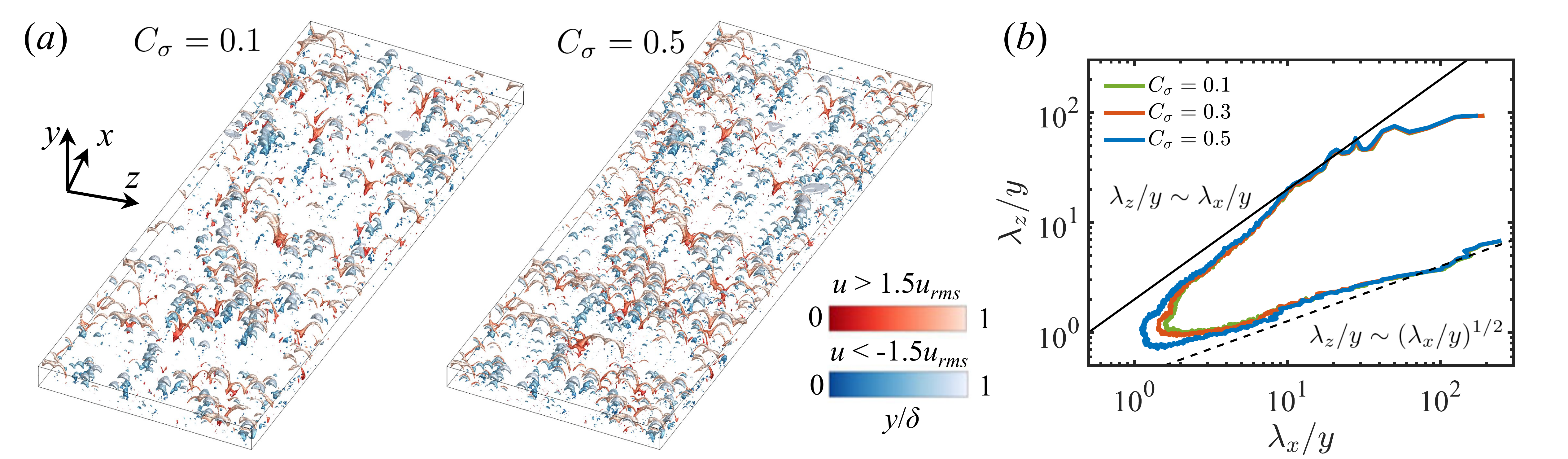}
	\caption{(a) Iso-surfaces of detached $u$ clusters in the SWAT at $\Rey_\tau=1000$ for $C_\sigma=0.1$ and $0.5$. (b) Two-dimensional spectra of the streamwise velocity at $y^+=100$ for $C_\sigma=0.1$, $0.3$ and $0.5$, with the contour level set to $k_x k_y E_{uu}/u_\tau^2 = 0.06$. The black solid and dashed lines denote the $ \lambda_z/y \sim \lambda_x/y$ and $\lambda_z/y \sim ( \lambda_x/y )^{1/2}$ relationships respectively.}
	\label{fig:csigma}
\end{figure}

The size and abundance of detached structures are influenced by the degree of vorticity concentration near the head of the hairpin vortex, which is controlled by the core variation coefficient $C_\sigma$. This parameter specifies the ratio of tube thickness between the head and the legs of the vortex. A smaller $C_\sigma$ indicates a weaker head–leg contrast, while a larger $C_\sigma$ corresponds to a more concentrated head region. Figure~\ref{fig:csigma}(a) supplements the comparison by showing iso-surfaces of detached $u$ clusters for $C_\sigma=0.1$ and $C_\sigma=0.5$. Together with figure~\ref{fig:ucluster}(c), corresponding to $C_\sigma=0.3$, these results reveal a systematic trend: as $C_\sigma$ increases, detached structures become increasingly abundant, indicating the role of vortex-core variation in regulating the attached–detached balance.

To quantify this trend, table~\ref{tab:unumber} reports the numbers and relative fractions, with respect to the total population, of all $u$ clusters, attached $u$ clusters, and detached $u$ clusters identified in SWAT at $\Rey=1000$ for three values of the core variation coefficient ($C_\sigma=0.1$, $0.3$, and $0.5$). As $C_\sigma$ grows, vorticity increasingly concentrates toward regions farther from the wall, causing the population of attached structures to gradually decrease. At the same time, detached structures not only increase in number but also account for a larger proportion of the total cluster population.

\begin{table*}
\caption{\label{tab:unumber} Number and relative fractions (with respect to the total population) of all $u$ clusters, attached $u$ clusters, and detached $u$ clusters identified in the SWAT at $\Rey_\tau=1000$ for different core variation coefficients $C_\sigma=0.1$, $0.3$, and $0.5$.} 
\begin{ruledtabular}
\setlength{\tabcolsep}{2.5mm}
\begin{tabular}{cccccc}
	$C_\sigma$ & $N_{\text{total}}$ & $N_{\text{attached}}$ & $N_{\text{detached}}$ & $N_{\text{attached}}/N_{\text{total}}$  & $N_{\text{detached}}/N_{\text{total}}$\\ \hline
	0.1 & 6581 & 1267 & 5314 & 0.19 & 0.81\\
	0.3 & 8338 & 951 & 7387 & 0.11 & 0.89\\
	0.5 & 12139 & 785 & 11354 & 0.06 & 0.94
\end{tabular}
\end{ruledtabular}
\end{table*}
	
The effect of $C_\sigma$ is also reflected in the spectral distribution. In the two-dimensional spectra, larger $C_\sigma$ values lead to a progressive broadening of the energy distribution, allowing more energy to populate smaller scales (see figure~\ref{fig:csigma}(b)). It is worth noting that in the SWAT framework, $C_\sigma$ has little influence on the low-order statistical quantities (see appendix~\ref{app:sensitivity}). Nevertheless, as discussed above, it exerts a pronounced influence on the relative balance between attached and detached structures, as well as on the distribution of small-scale energy. These results highlight that in real turbulence, the geometry of vortex structures and the associated vorticity distribution are not merely geometric details, but may provide a key physical mechanism governing the coexistence of attached and detached motions and their role in shaping the scale-by-scale energy distribution.

\subsection{Organization of large-scale motions}

A defining feature of wall turbulence lies not only in the geometry of individual vortical structures but also in their collective organization across scales. In this section, we investigate how the arrangement of vortex packets, their spanwise meandering, and the alignment of VLSMs govern the emergence of flow patterns and statistical signatures in wall turbulence. Leveraging the flexibility of SWAT, we systematically disentangle the roles of packet organization in producing realistic streaky velocity fields and long-range correlations, assess the influence of spanwise meandering in regulating streamwise energy levels, and demonstrate the ability to prescribe and control the orientation of VLSMs. These analyses establish the importance of structural organization in shaping turbulence dynamics and highlight the utility of SWAT as a framework for probing the multi-scale coherence of wall-attached motions.

\begin{figure}
	\centering
	\includegraphics[width=0.95\linewidth]{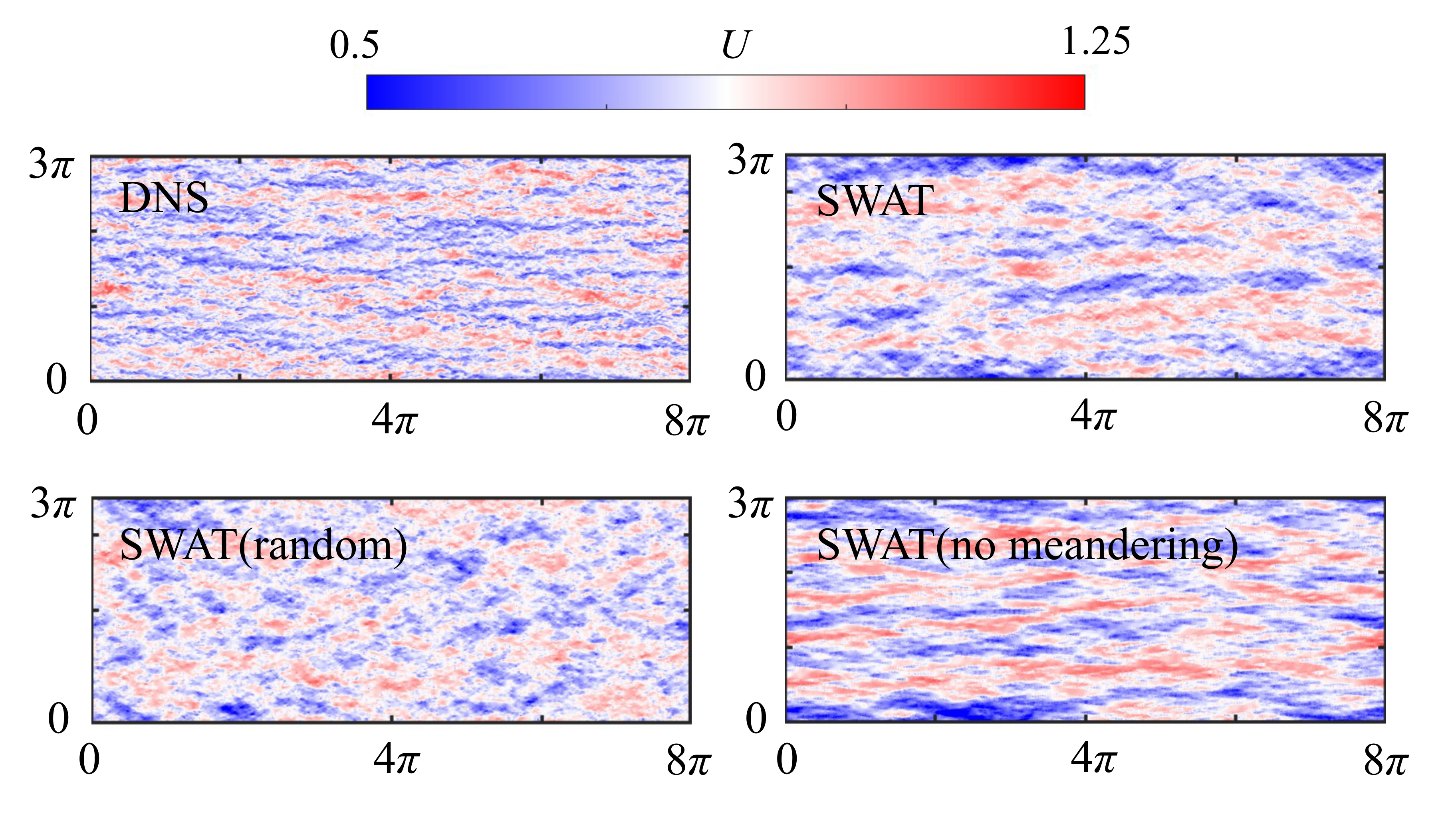}
	\caption{Streamwise velocity contours from DNS and SWAT at the center of the logarithmic region $y^+=3.9Re_\tau^{1/2}$ for $\Rey_\tau=1000$: (a) DNS; (b) the standard SWAT with both packet arrangement and meandering; (c) SWAT without packet arrangement (random); (d) SWAT with packet arrangement but without meandering.}
	\label{fig:contours}
\end{figure}

To assess the role of packet arrangement and spanwise meandering, we first analyze streamwise velocity contours at the centre of the logarithmic region, $y^+ = 3.9Re_\tau^{1/2}$, comparing DNS data with SWAT cases both with and without vortex packet arrangement and spanwise meandering (see figure~\ref{fig:contours}). The comparison shows that the structured arrangement of vortex packets combined with spanwise meandering successfully reconstructs the streaky patterns characteristic of streamwise velocity fields in wall turbulence, closely resembling the DNS results \citep{Hutchins2007Evidence}. These streaks, consisting of alternating low- and high-speed regions, are canonical coherent structures that play a central role in momentum transport and turbulence dynamics. By contrast, a random vortex arrangement generates irregular, patchy velocity structures that fail to reproduce these essential features, thereby highlighting the importance of organized vortex packets in capturing the physics of wall turbulence. For the SWAT case with packet arrangement but without meandering, streaky patterns are still evident. However, the overly aligned vortex structures cause the velocity disturbances to be excessively concentrated, which manifests as unnaturally intense streaks in the contour plots.

\begin{figure}
	\centering
	\includegraphics[width=\linewidth]{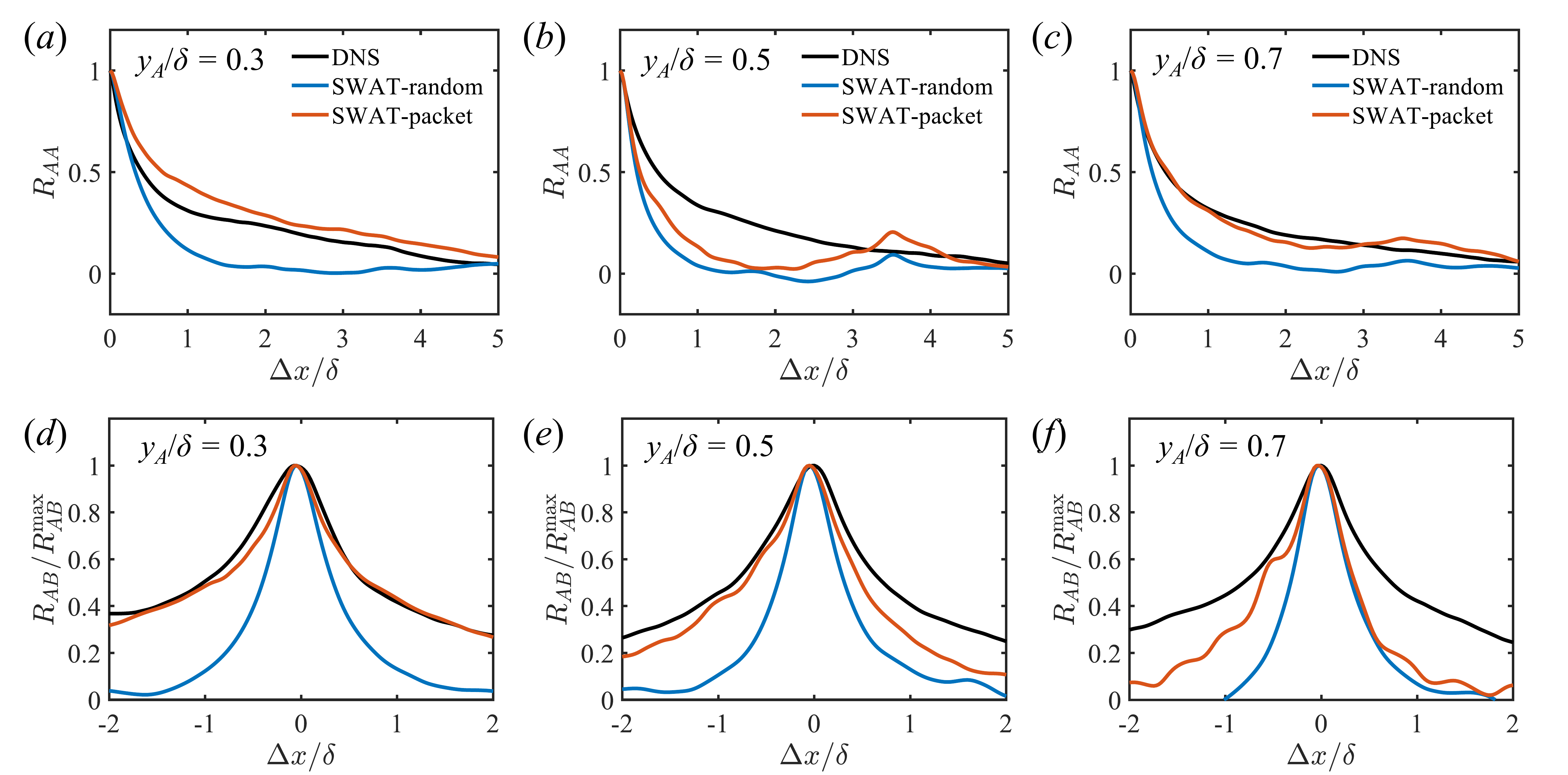}
	\caption{(a-c) Autocorrelation of the streamwise velocity $u$ at $y_A$ and (d-f) normalized two-point cross-correlation of $u$ between $y_A$ and $y_B=y_A+\Delta y$, evaluated at different wall-normal positions through the boundary layer: (a,d) $y_A/\delta=0.3$, (b,e) $0.5$, and (c,f) $0.7$. Black lines denote DNS data, blue lines denote SWAT without packet arrangement (random), and red lines denote the standard SWAT (packet). For the two-point statistics, the wall-normal spacing is $\Delta y/\delta=0.1$.}
	\label{fig:correlation}
\end{figure}

The organization of vortex packets also has a pronounced impact on the streamwise velocity correlations \citep{Marusic2001On}. Figure~\ref{fig:correlation} compares DNS data with SWAT cases both with and without packet arrangement, examining two measures of the streamwise velocity $u$: the autocorrelation at a single height and the normalized two-point cross-correlation between two wall-normal positions. Both statistics are evaluated at different wall-normal positions across the boundary layer. For a completely random arrangement of hairpin vortices, both the single-point autocorrelation and the two-point cross-correlation decay rapidly with increasing spatial separation, falling below the levels observed in the DNS data. In contrast, the introduction of organized vortex packets significantly enhances streamwise correlations, consistent with the mechanism proposed by~\citet{Marusic2001On}, where the alignment of hairpin packets sustains long streamwise coherence and yields correlation levels closely matching the DNS results.

\begin{figure}
	\centering
	\includegraphics[width=0.9\linewidth]{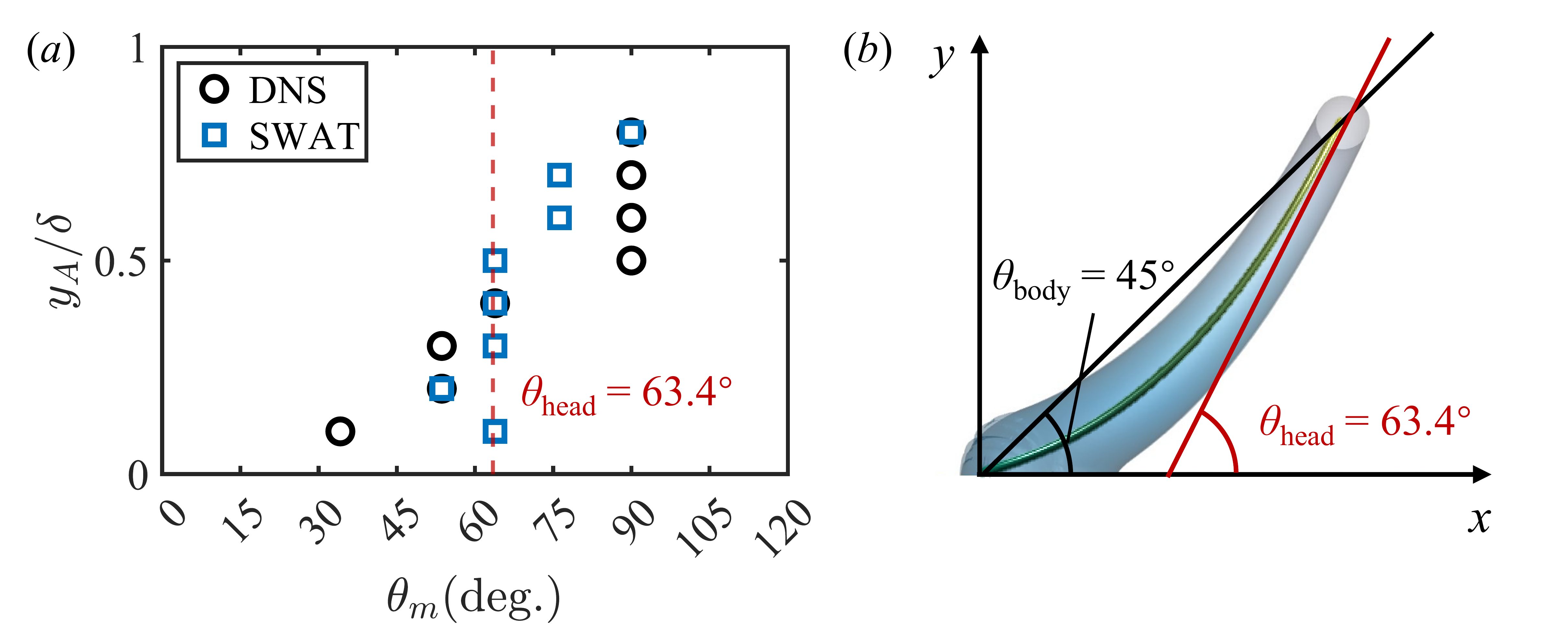}
	\caption{(a) Streamwise inclination angles from DNS and from SWAT constructed with hairpin vortices of body inclination $\theta_\text{body}=45^\circ$ and head inclination $\theta_\text{head}=63.4^\circ$. (b) Side view of the hairpin-vortex building block.}
	\label{fig:angle}
\end{figure}

The two-point cross-correlation measurements enable us to estimate the streamwise inclination angle of large wall-attached structures in turbulent boundary layers. Figure~\ref{fig:angle}(a) shows the inclination angles computed from DNS and from SWAT at different wall-normal positions. Previous studies \citep{Deshpande2019Streamwise,Cheng2022Streamwise} have reported that the inclination angle of attached eddies is close to $45^\circ$. In our SWAT design, the hairpin vortices were likewise prescribed with a body inclination of $\theta_\text{body}=45^\circ$. However, the angles obtained from the cross-correlation analysis concentrate instead around $63.4^\circ$, notably larger than the prescribed body inclination. This discrepancy arises because the lateral legs of the hairpin are curved, such that the local inclination varies along the streamwise direction, and the correlation diagnostic preferentially captures the portion with the strongest coherent imprint.

Importantly, the analysis reveals that correlation-based inclination estimates are not dictated by the nominal body inclination but are systematically governed by the vorticity-concentrated head of the hairpin vortex, which, in our configuration, has $\theta_\text{head}=63.4^\circ$ (see figure~\ref{fig:angle}(b)). This head-dominated response explains the higher measured angles and provides a new interpretation of previous inclination-angle measurements in real flows. More broadly, it highlights the need for an explicit head–body distinction in attached-eddy modelling: the correlation signatures that underpin experimental and numerical diagnostics primarily reflect the head geometry and its induced velocity field, rather than the nominal body inclination. Such a distinction is absent in the classical AEM and may help reconcile discrepancies in inclination-angle reports. This insight therefore adds a critical refinement to the interpretation of inclination angles and points to new directions for improving the AEM framework, where future refinements should explicitly parameterize or constrain head geometry and strength in addition to the body tilt.

\begin{figure}
	\centering
	\includegraphics[width=0.9\linewidth]{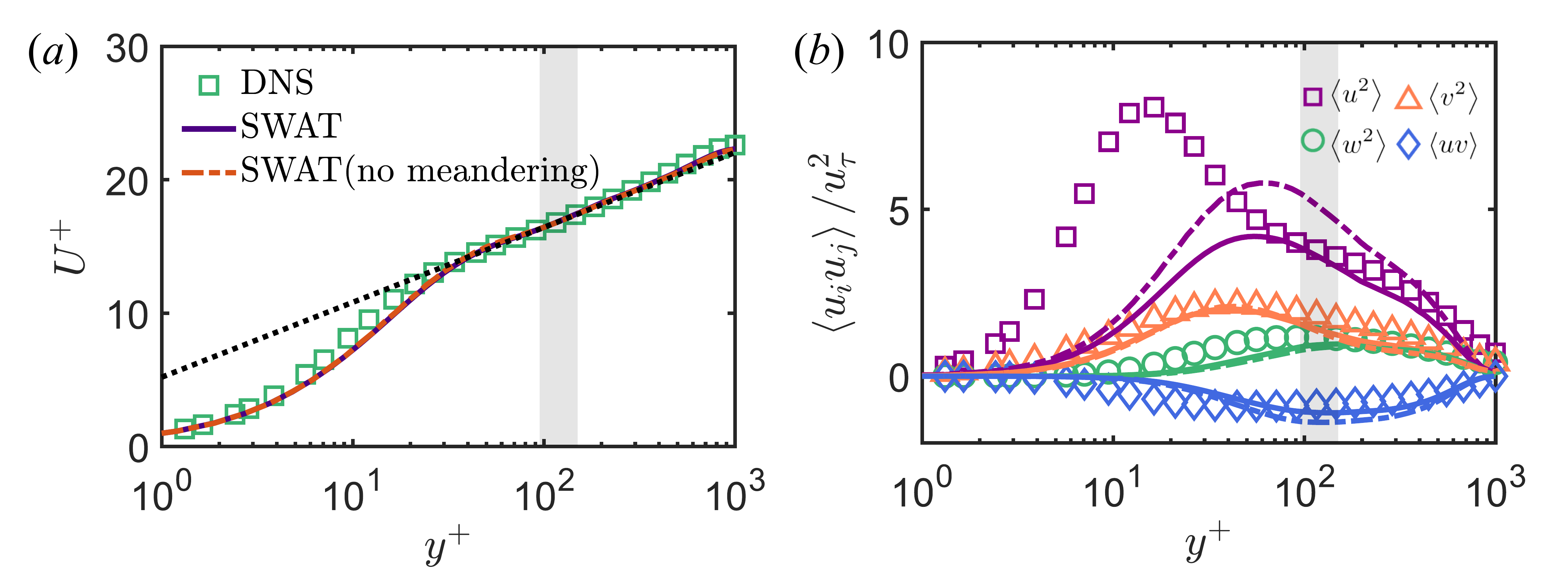}
	\caption{Comparison of (a) mean velocity profiles and (b) Reynolds stress profiles between SWAT with spanwise meandering (solid lines), SWAT without spanwise meandering (dash–dotted lines), and DNS data (symbols) at $\Rey_\tau = 1000$.}
	\label{fig:meanderingsta}
\end{figure}

Figure~\ref{fig:meanderingsta} compares the mean velocity profiles and Reynolds stress components between DNS data and SWAT cases with and without spanwise meandering. The inclusion of spanwise meandering in the vortex packets has little effect on the mean velocity profile, as well as on the Reynolds stress components $\langle v^2 \rangle$, $\langle w^2 \rangle$, and $\langle uv \rangle$. In contrast, it strongly affects the streamwise Reynolds stress component $\langle u^2 \rangle$: in the absence of meandering, the synthesized wall turbulence exhibits significantly higher $\langle u^2 \rangle$ levels than the meandering case (figure~\ref{fig:meanderingsta}(b)). This demonstrates that spanwise meandering plays a crucial role in regulating the intensity of streamwise fluctuations, thereby preventing excessive energy concentration in the streaks. This mechanism is also evident from the velocity contour comparisons shown in figure~\ref{fig:contours}.

\begin{figure}
	\centering
	\includegraphics[width=\linewidth]{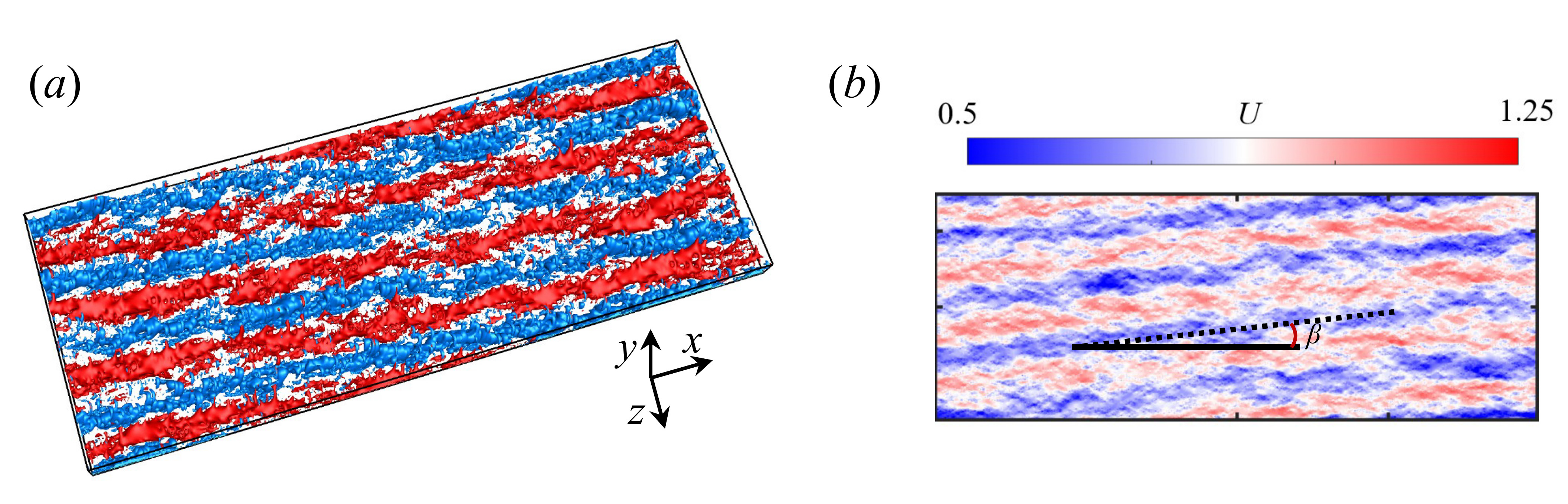}
	\caption{Spatial organization of VLSMs designed in SWAT at $\Rey_\tau = 1000$ with a prescribed characteristic angle $\beta = 6^\circ$. (a) Iso-surfaces of the streamwise velocity, where the red and blue surfaces correspond to $u/U_\text{bulk}=0.1$ and $-0.1$, respectively, highlighting the presence of VLSMs. (b) Streamwise velocity contours at the center of the logarithmic region, $y^+=3.9Re_\tau^{1/2}$, showing the alignment of VLSMs at the characteristic angle $\beta = 6^\circ$ relative to the streamwise direction.}
	\label{fig:VLSM}
\end{figure}

We next turn to the organization of VLSMs. Taking advantage of the flexibility of SWAT in prescribing structural arrangements, we consider the spatial alignment of VLSMs. Previous studies \citep{Baltzer2013Structural, Lee2019Space} have shown that VLSMs exhibit alignment behaviour, often orienting along dominant roll-cell motions at characteristic angles relative to the streamwise direction. In particular, pipe-flow studies have reported characteristic VLSM angles of approximately $4$–$8^\circ$ with respect to the streamwise axis. 
While traditional AEMs describe the statistical organization of wall-attached motions, they do not explicitly resolve or control the geometric alignment of VLSMs. Here, we explicitly demonstrate that such alignment can be faithfully generated and tuned within SWAT. 
To this end, we constructed a SWAT configuration in which only the wall-coherent superstructures, corresponding to the largest-scale vortex packets (see figure~\ref{fig:packet}(c)), were arranged with a prescribed inclination angle of $\beta = 6^\circ$, while the smaller-scale hierarchy remained randomized. Figure~\ref{fig:VLSM}(a) presents well-organized iso-surfaces of streamwise velocity fluctuations. These structures align with the designed inclination.
Moreover, the measured angles from the streamwise velocity contours in figure~\ref{fig:VLSM}(b) closely match the prescribed superstructure arrangement, confirming the ability of SWAT to reproduce and control the characteristic orientation of VLSMs. This alignment originates solely from the imposed organization of the superstructures, leaving the self-similar hierarchy statistically unaltered and preserving all similarity-based scaling properties.

\subsection{Contributions of individual hierarchies}\label{app:contributions} % to statistical profiles

As the fundamental building blocks of the SWAT model, vortex packets across different hierarchies and the superstructure collectively contribute to a realistic approximation of the statistical profile. Here, we analyze the influence of vortex packets at different hierarchies and superstructures on the overall statistical properties. Specifically, we quantify how individual hierarchical contributions shape the mean velocity profile using the velocity defect
\begin{equation}
	\tilde{U}_{d,i} = \tilde{U}_{o,i} - \left\langle \tilde{U}_i\right\rangle,
\end{equation} 
where $\tilde{U}_{o,i} = \left\langle \tilde{U}_i \right\rangle_{y=\delta}$ represents the centerline velocity contributed by the $i$th self-similar hierarchy.  
The total induced velocity $\tilde{U}$, together with the corresponding centerline velocity $\tilde{U}_o$, can be reconstructed through the principle of superposition
\begin{equation}
	\tilde{U} = \sum_{i=1}^{N_p} \tilde{U}_{i} + \tilde{U}_{SS}
	,\quad 
	\tilde{U}_o = \sum_{i=1}^{N_p} \tilde{U}_{o,i} + \tilde{U}_{SS,o},
\end{equation}  
where $\tilde{U}_{SS}$ represents the contribution from the superstructure hierarchy.  
With a uniform bulk velocity, $\bar{U}$, and the implementation of the damping function \eqref{eq:damping}, the turbulent velocity profile $U$ is obtained.  
This formulation naturally recovers the classical velocity defect law
\begin{equation}\label{eq:defect}
	\frac{U_o - \langle U \rangle}{u_\tau} = -\kappa^{-1} \ln\left(\frac{y}{\delta}\right) + B_1, \
\end{equation}  
with the von Kármán constant $\kappa=0.41$ and constant $ B_1 = 0.2 $ \citep{Pope2000Turbulent}. 

\begin{figure}
	\centering
	\includegraphics[width=0.9\linewidth]{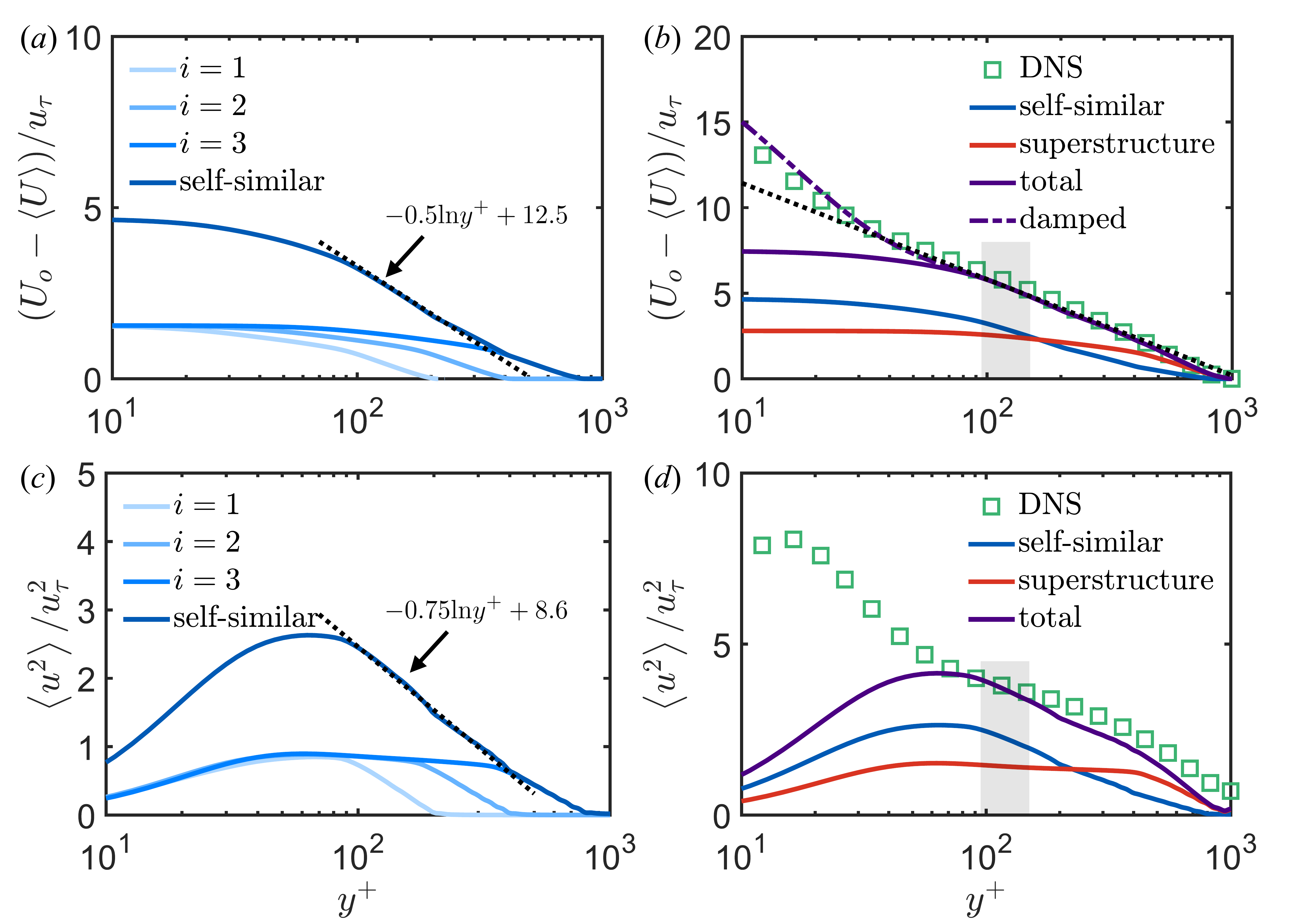}
	\caption{Contributions of self-similar hierarchies and the superstructure hierarchy to the (a, b) mean velocity profile and (c, d) Reynolds stress profile. Panels (a) and (c) show individual contributions from the three self-similar hierarchies, along with their cumulative effect representing the self-similar component. Panels (b) and (d) include the additional contribution from the superstructure hierarchy and the total combined profile. The shaded grey region indicates the log-law region. The dotted line in (b) denotes the velocity defect law \eqref{eq:defect}.}
	\label{fig:contribution}
\end{figure}

To illustrate these contributions, we consider a representative SWAT at $ \Rey_\tau = 1000 $. Figure~\ref{fig:contribution}(a) separately displays the contributions to statistics from three self-similar hierarchies and their aggregated effect, which collectively captures the logarithmic behaviour within the log region. Subsequently, the incorporation of the superstructure hierarchy yields a velocity defect profile that closely corresponds with the DNS data in the outer region, as illustrated in figure~\ref{fig:contribution}(b). The cumulative summation of these components closely reconstructs the velocity defect profile, confirming the additive nature of these structures in shaping the mean flow. 
An analogous decomposition is evident for the Reynolds stress profiles, as depicted in figure~\ref{fig:contribution}(c).  
Here, the three self-similar hierarchies primarily recover the logarithmic region, while the inclusion of the superstructure hierarchy is essential for accurately reproducing the full Reynolds stress profile, especially in the outer region, as shown in figure~\ref{fig:contribution}(d). In the present SWAT model, superstructures are prescribed up to the boundary-layer thickness $\delta$. 
Nevertheless, their contribution to outer-layer statistics is strongly diminished for $y/\delta \gtrsim 0.5$, consistent with numerical and experimental observations \citep{Baars2017Self,Deshpande2020Two}.

\section{Application of SWAT in initializing simulations of wall turbulence}\label{subsec:CF}
Due to the statistical and structural consistency of our synthetic turbulence with real turbulence, it provides an efficient and cost-effective alternative for initializing numerical simulations. This approach eliminates the need for extensive computational resources otherwise required to wait for the flow to transition and fully develop. This advantage is particularly beneficial in high-fidelity simulations such as DNS of channel flow, where computational resources are often a limiting factor.  

To generate synthetic channel turbulence, hierarchical hairpin vortex packets are independently constructed on both the upper and lower walls. For the channel flow configuration, the boundary layer height is set to half the channel height, $\delta = L_y/2$. Hairpin vortices of varying hierarchical levels are introduced at each wall, preserving essential features such as streamwise alignment and spanwise meandering within vortex packets. To maintain statistical realism, the spatial distribution of these vortex packets is randomized independently for the top and bottom walls. As described in \S\,~\ref{subsec:construction}, the damping function~\eqref{eq:damping} is applied separately to each half of the channel, enforcing the no-slip boundary condition and ensuring that the synthetic velocity field is suitable as initial conditions for DNS. 

The DNS is performed by solving the 3D incompressible Navier-Stokes equations using a Fourier-Chebyshev pseudo-spectral method~\citep{Kim1987Turbulence}.  The computational domain dimensions are $L_x = 2\pi$ in the streamwise direction, $L_y = 2\delta$ in the wall-normal direction (with $\delta = 1$), and $L_z = \pi$ in the spanwise direction. 
Note that though larger domain is usually used for practical channel flow simulations, the present dimensions are sufficiently large to capture accurate one-point flow statistics~\citep{LozanoDuran2014Effect}, which is the purpose of the current tests.
The simulations are conducted at a friction Reynolds number of $\Rey_\tau = 1000$ and a bulk Reynolds number of $\Rey \approx 20000$, using a grid resolution of $N_x \times N_y \times N_z = 768 \times 769 \times 768$ \citep{Wang2022Synchronization}. The flow is driven by a constant pressure gradient in the streamwise direction.

\begin{figure}
	\centering
	\includegraphics[width=0.7\linewidth]{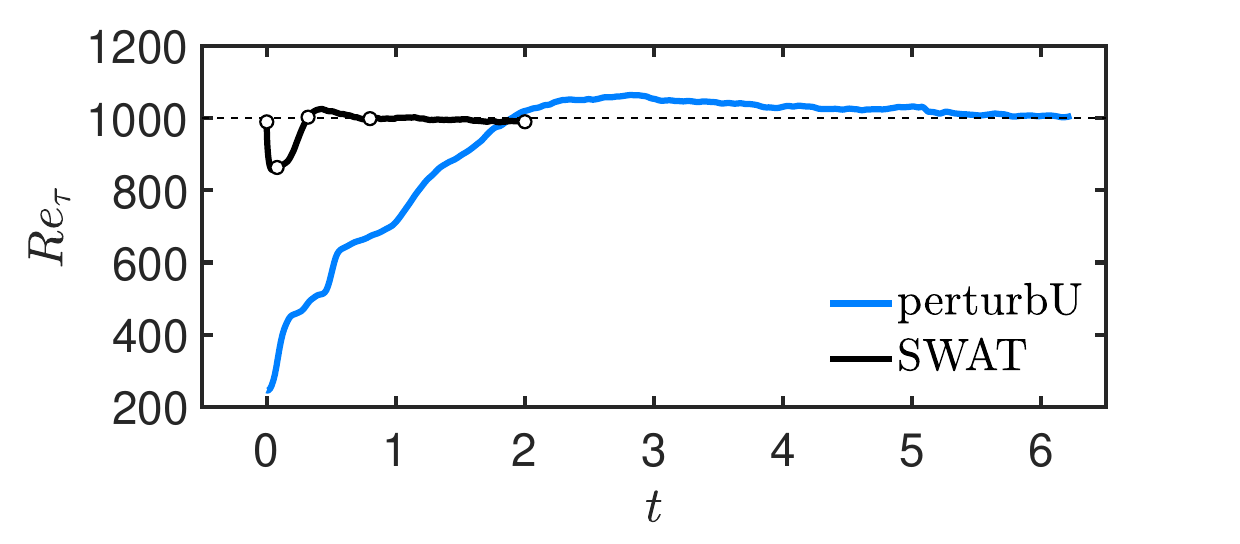}
	\caption{Time evolution of the friction Reynolds number in DNS simulations with different initialization methods. The blue line corresponds to the traditional perturbation method (perturbU), while the black line represents SWAT. The flow field at the time indicated by the circle on the black line is shown in figure \ref{fig:swatdns}.}
	\label{fig:compare_iterations}
\end{figure}

To assess the computational efficiency of the synthetic turbulence method, we compare it to the traditional perturbation initialization ‘perturbU’~\citep{de2006potential}, which introduces perturbations by superimposing a fluctuating velocity field, $\boldsymbol{u}$, onto the Poiseuille base flow, $U_0$. This approach mimics key aspects of the near-wall turbulence cycle, allowing for the transition to turbulence to occur relatively quickly. However, the effectiveness of this method depends on the gradual amplification of small perturbations, which can result in a longer overall transition time.  

To quantify the computational cost of achieving fully developed turbulence, we compare the time-dependent evolution of the Reynolds number in DNSs using both initialization methods (see figure~\ref{fig:compare_iterations}). The perturbU approach initiates turbulence through the progressive strengthening of initial vortices, which subsequently undergo asymmetric breakdown, enhancing turbulent mixing and promoting vortex formation~\citep{de2006potential}. As turbulence production intensifies, momentum transfer increases, gradually eroding the parabolic velocity profile. Eventually, the flow stabilizes into a statistically steady state characterized by a logarithmic velocity distribution. Despite effectively triggering turbulence, this process is computationally expensive, requiring more than five flow-through times (FTTs) over 129,000 CPU hours for turbulence to fully develop at the target $\Rey_\tau$ due to the slow evolution of the flow structures.

\begin{figure}
	\centering
	\includegraphics[width=\linewidth]{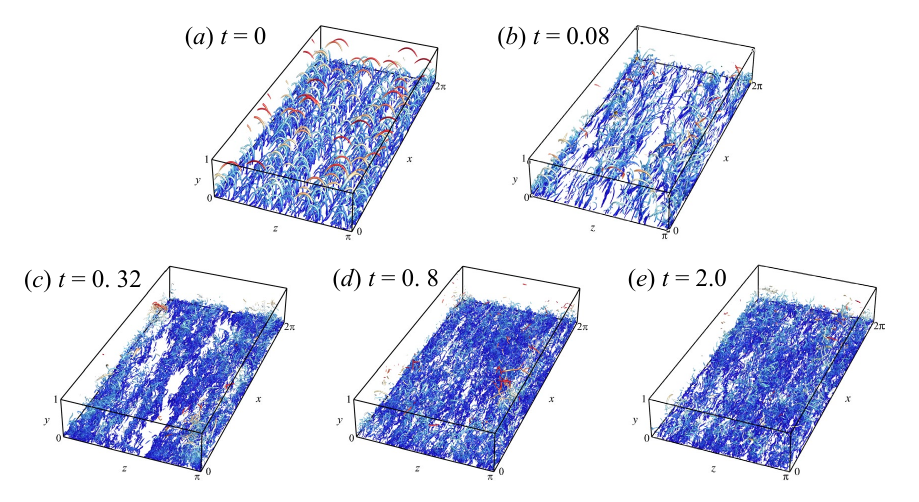}
	\caption{Temporal evolution of the isosurface of the swirling strength \citep{Zhou1999Mechanisms} $\lambda_{ci} = 0.1 \lambda_{ci,\max}$ depicting vortex structures in a DNS of turbulent channel flow initialized with the SWAT field at $\Rey_\tau = 1000$, where $\lambda_{ci,\max}$ denotes maximum swirling strength $\lambda_{ci}$ value in the flow field at a given time instant. The friction Reynolds number $\Rey_\tau$ corresponding to each flow field is annotated in the circles shown in figure~\ref{fig:compare_iterations}. }
	\label{fig:swatdns}
\end{figure}

In contrast, the SWAT method for channel flow generates a well-structured turbulence field from the beginning, featuring meandering low- and high-speed streaks alongside densely packed hairpin vortex bundles. These structures introduce an appropriate distribution of turbulent kinetic energy, ensuring that both turbulence characteristics and statistics closely align with the target Reynolds number early in the evolution. As the flow progresses, interactions between vortices near the upper and lower walls further enhance the transition to fully developed turbulence (see figure~\ref{fig:swatdns}). This transition occurs rapidly and efficiently, preserving the essential turbulence structures while minimizing computational overhead. Although there is a slight initial drop in the Reynolds number, it quickly recovers to the target value within $O(1)$ FTTs (requiring only 35,800 CPU hours). Furthermore, due to the well-matched initial statistics provided by SWAT, the mean velocity profiles and Reynolds stresses converge to their true equilibrium states significantly faster than with the perturbU approach. By contrast, the traditional perturbation method requires considerably longer to reach the same level of turbulence development, making SWAT a more efficient alternative for simulations requiring rapid and reliable turbulence initialization.

Beyond channel flow, SWAT can be integrated into other types of wall-bounded turbulence, such as flat plate turbulent boundary layers and wall turbulence over complex curved surfaces. For example, in the case of a turbulent boundary layer, where the boundary layer thickness increases downstream, a SWAT model scaled with the local normalized boundary layer thickness can adjust the size and height of hairpin vortex packets at different physical locations to match the local flow physics. For wall turbulence over curved surfaces, the hairpin vortex structures originally generated in a Cartesian grid can be mapped onto the corresponding curvilinear coordinates.

Furthermore, the constructed SWAT flow fields can potentially serve as inflow conditions for spatially developing wall-bounded turbulence simulations. In this context, the synthetic turbulence generated by SWAT can be prescribed at the inflow plane of a computational domain to provide a time-varying velocity field at the inlet boundary. Previous studies based on the AEM approach \citep{Subbareddy2006A} have shown that such inflow conditions are both cost-effective and promising. Nevertheless, further detailed investigations are necessary to evaluate the accuracy and fidelity of the downstream turbulence evolution when such synthetic inflow methods are employed.

\section{Conclusions}\label{subsec:discussion}

This study presents a systematic approach to constructing wall turbulence using hierarchically organized hairpin vortex packets. The method is carefully designed and well encapsulated, involving only a few fixed parameters without empirical tuning.
It enables the efficient generation of SWAT at a range of given Reynolds number while faithfully reproducing the coherent structures and key statistical characteristics observed in wall-bounded turbulent flows. Beyond supporting existing AEM theories, the approach also provides new insights into vortex configuration and organization for advancing wall-turbulence and AEM research, as well as practical utility for numerical simulation initialization.

Combining insights from numerical simulations and experimental observations, we employ realistic hairpin vortices with geometrically complex centerlines and vortex core sizes that vary along their centerlines as the fundamental building blocks of wall turbulence. Hairpin vortices of different heights are then organized into vortex packets with spanwise meandering structures. Using the attached-eddy model, these hierarchically structured hairpin vortex packets self-similarly populate the wall with different population densities at varying scales and levels. Additionally, wall-coherent superstructures within the largest vortex packets are incorporated as VLSM to supplement the outer layer dynamics. Finally, by carefully integrating inflow conditions and boundary constraints, the method generates synthetic wall turbulence at any prescribed Reynolds number with minimal computing resources.

Through direct comparisons with DNS data, the SWAT model successfully replicates key statistical and structural features of turbulent channel flows across different Reynolds numbers. It accurately captures the mean velocity profile, including the scaling in the logarithmic region, without relying on external flow datasets. The log-law naturally emerges from the collective behaviour of hairpin vortices, while the transitions in the viscous sublayer and buffer layer are self-consistent. Moreover, the model successfully reproduces the anisotropic distribution of Reynolds stresses while preserving the correct scaling laws in the logarithmic region.  
Furthermore, the streamwise energy spectra obtained from SWAT reproduce the expected $k_x^{-1}$ scaling at large scales and its Reynolds-number-dependent extension, with plateau values consistent with reported estimates of the Townsend–Perry constant $A_1$, in good agreement with DNS results and theoretical AEH predictions. Moreover, SWAT demonstrates the ability to capture higher-order statistics, such as even-order velocity moments. While the model successfully reproduces many key features, some discrepancies suggest avenues for further improvement, such as refining the representation of inner-outer interaction dynamics~\citep{Marusic2010Predictive} to enhance the accuracy of large-scale motions and their coupling with near-wall structures.  

The insights gained from SWAT highlight the central role of vortex geometry, packet organization, and hierarchical contributions in shaping wall turbulence. The improved hairpin-vortex model reproduces both attached and detached motions, with the relative abundance of detached motions controlled by the vortex-core variation, thereby linking structural geometry to energy distribution across scales. In contrast to earlier AEM-based approaches that required ad hoc detached vortices, this replenishment is achieved naturally through the height-dependent core-size variation. In terms of vortex packet organization, the inclusion of spanwise meandering and hierarchical vortex arrangements enables the model to effectively capture the large-scale streaky patterns of streamwise velocity, yielding realistic correlation statistics and inclination angles consistent with observations, while regulating streamwise energy intensity. Importantly, the analysis further reveals that correlation-based inclination estimates are systematically governed by the vorticity-concentrated head of the hairpin vortices rather than by the prescribed body inclination. Furthermore, the ability to prescribe very-large-scale orientations demonstrates that SWAT can replicate and control the alignment of VLSMs. The decomposition of statistical contributions confirms that self-similar hierarchies primarily recover the logarithmic region, whereas the inclusion of superstructures is crucial for matching outer-layer behaviour, underscoring the additive and multi-hierarchical nature of wall turbulence.

Beyond its physical accuracy, our method provides a practical advantage by enabling the rapid generation of initial conditions for wall turbulence at any Reynolds number within arbitrary computational domains without requiring additional data. This capability significantly reduces computational costs in the early stages of simulations, whether for DNS or LES, while ensuring statistical and structural consistency with real turbulence.  

Notably, our stance is methodological rather than morphological. 
Whether canonical hairpin vortices persist at high Reynolds numbers remains an open question, as their signatures in instantaneous fields lose visual coherence with increasing $\Rey$ and thereby become challenging to identify \citep{Jiménez2018Coherent}.
We do not intend to establish hairpin vortices as the sole or universal structure, nor to expect them to reproduce all statistics and features at high Reynolds numbers. Instead, we present SWAT as a testable, parameter-controlled conceptual platform that links structural geometry, organizational patterns, and statistical responses along a causal chain. The framework is agnostic to the choice of primitive: in future applications and validation, hairpin vortices could be replaced by inclined shear layers or roller-like elements. At the same time, refining the hairpin design, including geometry, core parametrization, and packet organization, together with a sharper representation of inner–outer interactions, offers clear opportunities to further enhance fidelity across scales.

Looking ahead, the framework can be extended to more complex geometries and rough surfaces, and will next be assessed in other canonical wall-bounded flows, such as transitional boundary layers and pipe flow, which provide critical benchmarks for accuracy and robustness.
In addition, the ability to rapidly generate turbulence conditions makes it well suited for integration into adaptive and dynamic simulation environments, with potential benefits for aircraft aerodynamics, weather modeling, and industrial flows, where turbulence conditions frequently evolve.

\appendix

\section{Sensitivity analysis of model parameters for synthetic wall turbulence}\label{app:sensitivity}

In this appendix, we analyze the sensitivity of the constant parameters introduced in §\ref{sec:method} for SWAT, specifically the core variation coefficient $ C_\sigma $ in \eqref{eq:sigmaij} and the circulation coefficient $ C_\Gamma$ in \eqref{eq:Gammaij}.

\begin{figure}
 \centering
 \includegraphics[width=0.95\linewidth]{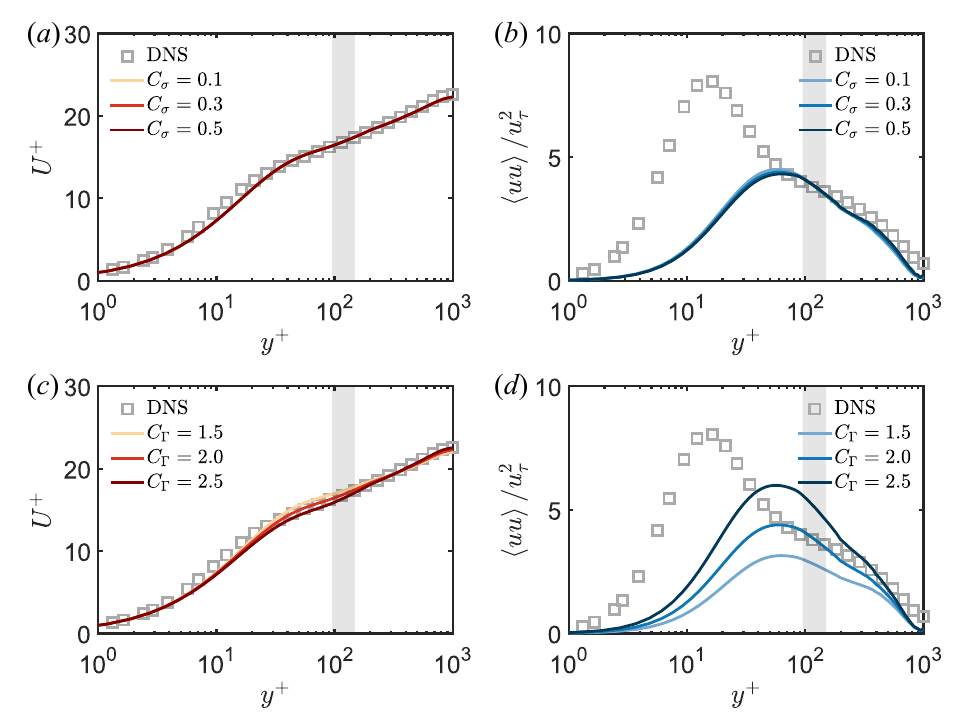}
 \caption{Sensitivity analysis of (a, c) the mean velocity and (b, d) Reynolds stress profile to variations in (a, b) the core variation coefficient ($C_\sigma=0.1,\,0.3,\,0.5$) and (c, d) the circulation coefficient ($C_\Gamma=1.5,\,2.0,\,2.5$). Reference data from DNS at $Re_\tau = 1000$ are shown as squares (mean velocity) and circles (Reynolds stresses). Corresponding predictions from the SWAT model are represented by red lines (mean velocity) and blue lines (Reynolds stresses).  }
 \label{fig:sensitivity}
\end{figure}

Figures~\ref{fig:sensitivity}(a,b) illustrate the sensitivity of the mean velocity profile and Reynolds stresses to variations in the core variation coefficient, tested at $ C_\sigma = 0.1,\,0.3,\,0.5 $. The results indicate that both the mean velocity and Reynolds stress profiles remain largely insensitive to changes in this parameter, and we select $ C_\sigma = 0.3 $ for our model. 

For the circulation coefficient, we perform a sensitivity analysis for $ C_\Gamma = 1.5,\,2.0,\,2.5 $, as shown in figures~\ref{fig:sensitivity}(c,d). Since the induced velocity of each hairpin vortex scales proportionally with the circulation coefficient, i.e., $ \Delta U_i^{(j)} \propto \Gamma_i^{(j)} \propto C_\Gamma $, variations in $ C_\Gamma $ primarily affect the Reynolds stress components, particularly the slope of $\langle u^2 \rangle$ in the logarithmic region. In contrast, the influence on the mean velocity profile is relatively minor.  Therefore, careful selection is necessary to accurately reproduce velocity moments in SWAT velocity fields. Based on these observations, we set $ C_\Gamma = 2.0 $, for the cases across different Reynolds numbers, ensuring an optimal match with the mean velocity and Reynolds stress profiles. 

\begin{acknowledgments}
	This work has been supported by the National Natural Science Foundation of China (Grant Nos.~12432010, 92152202, 12525201 and 12588201), and the New Cornerstone Science Foundation through the XPLORER Prize.
    Authors thank Z. Zhou and X. Zhu for helpful comments. Numerical simulations were carried out on the Tianhe-2A supercomputer in Guangzhou, China.
\end{acknowledgments}

\bibliography{bibfile}

\end{document}